\documentclass[twocolumn,amsmath,amssymb,floatfix,prb,footinbib,superscriptaddress]{revtex4}

\usepackage{amsmath,amssymb,bm,graphicx,url,epsf,color}

\usepackage{natbib}
\usepackage[english]{babel}
\usepackage{wasysym}
\usepackage{MnSymbol}
\usepackage{amsmath}
\usepackage{graphicx}
\usepackage{float}
\usepackage{subfig}
\usepackage{pdfpages}
\usepackage{mathrsfs}
\usepackage[colorinlistoftodos]{todonotes}
\usepackage[colorlinks=true, allcolors=blue]{hyperref}

\makeatletter
\AtBeginDocument{\let\LS@rot\@undefined}
\makeatother

\makeatletter
\def\NAT@spacechar{}
\makeatother

\graphicspath{{figure/}}

\begin{document}
\title{Hybrid interacting quantum Hall thermal machine}

\author{S. Finocchiaro}
  \affiliation{Center for Nonlinear and Complex Systems, Dipartimento di Scienza e Alta Tecnologia, Universit\`a degli Studi dell'Insubria, Via Valleggio 11, 22100 Como, Italy} 
    \affiliation{Istituto Nazionale di Fisica Nucleare, Sezione di Milano, Via Celoria 16, 20133 Milano, Italy}

\author{D. Ferraro}
\affiliation{
Dipartimento di Fisica, Universit\`a di Genova, Via Dodecaneso 33, 16146, Genova, Italy
}
\affiliation{
SPIN-CNR, Via Dodecaneso 33, 16146, Genova, Italy
}
\author{M. Sassetti}
\affiliation{
Dipartimento di Fisica, Universit\`a di Genova, Via Dodecaneso 33, 16146, Genova, Italy
}
\affiliation{
SPIN-CNR, Via Dodecaneso 33, 16146, Genova, Italy
}

\author{G. Benenti}
  \affiliation{Center for Nonlinear and Complex Systems, Dipartimento di Scienza e Alta Tecnologia, Universit\`a degli Studi dell'Insubria, Via Valleggio 11, 22100 Como, Italy} 
    \affiliation{Istituto Nazionale di Fisica Nucleare, Sezione di Milano, Via Celoria 16, 20133 Milano, Italy}

\date{\today}

\begin{abstract}
We investigate a hybrid thermal machine based on a single closed quantum Hall edge channel forming a quantum dot. It is tunneling coupled with two quantum Hall states at $\nu=2$ in contact with reservoirs at different temperatures and chemical potentials. One of these edge states is also driven out-of-equilibrium by means of a periodic train of Lorentzian voltage pulses. This device allows to explore various possible working regimes including the engine, the heat pump and the refrigerator configuration. 
Regions where two regimes coexist can also be identified. 
Moreover, the proposed set-up exhibits robustness and in some parameter regions also slightly enhanced performance in the presence of electron-electron interactions.
\end{abstract}

\maketitle

\section{Introduction}

Since the first industrial revolution, the possibility to efficiently convert heat into useful work played a central role in boosting the technological development of thermal machines. The study of the physical principles behind this transformation led to the emergence of thermodynamics~\cite{Carnot,Fermi}.

Recent advances in miniaturization have triggered interest in devices able to deliver power, cooling, or heating at the microscopic level, paving the way for the rapidly growing field of quantum thermodynamics~\cite{Esposito09, Campisi11, Kosloff13, Kurizki15, Sothmann15, Campisi16, Vinjanampathy, Goold16,Benenti17, Pekola21, Landi21, Landi22, Potts24, Deffner_book}.
Indeed, the study of thermodynamics for microscopic and mesoscopic devices out-of-equilibrium actually requires a by far non-trivial extension of conventional thermodynamics towards the quantum realm. 
This is due to the fact that
in nanodevices genuine non-classical effects play a crucial role, and fundamental concepts such as heat and work need to be properly reconsidered.

A deep understanding of these basic concepts at the quantum level represents therefore a fundamental starting point for designing highly performant quantum thermal machines~\cite{Bhattacharjee21,Arrachea23,Cangemi2020,Cavaliere23, Cavaliere25}.
A first principles approach to the problem, starting from the non-classical properties that underlie the phenomenological laws of heat and particle transport, has proven particularly suitable for studying systems at the mesoscopic scale~\cite{Hicks93,Hicks,Hicks96}, demonstrating the potential to enhance the thermoelectric figure of merit in these systems. 
In this context, theoretical work has led to the possibility of controlling heat currents and designing thermal diodes, transistors, and thermal logic gates~\cite{Terraneo,Li}. 

Among the main questions emerging in this field of research is the possibility to realize \textit{hybrid} miniaturized  thermal machines able to perform multiple tasks, showing for example an heat pump behavior 
 with simulaneous production of useful work~\cite{Manzano20, Cavaliere23}. Great part of the studies about hybrid thermal machines consider stationary conditions and require multi-terminal geometries. However, driving the system and/or the system-bath couplings offers the possibility to switch among different thermodynamic tasks simply by tuning the driving frequency even in a simpler two-terminal geometry~\cite{Ludovico16}.

In this paper, we will show that such hybrid behavior can be realized with a closed quantum Hall channel tunneling coupled to two other quantum Hall edge channels connected to reservoirs kept at different temperatures and chemical potentials. In addition, one of these terminals is further driven out-of-equilibrium by a periodic voltage~\cite{Ryu22}. In this configuration, depending on the working point, such thermal machine can work as engine, heat pump and refrigerator. 
The engine (given by a particle current flowing agains its ``natural direction'' imposed by the chemical potential) and the heat pump can also occur together, performing then \textit{multiple tasks}. Moreover, an interesting but still largely unexplored aspect is the proper characterization of the role played by interaction and quantum correlations in modifying the efficiency of a quantum thermal machine~\cite{Benenti13, Braggio24}. In this direction, we will show that the presented device is also the ideal playground to study interaction effects in low dimension~\cite{Bocquillon14, Ferraro24}. Indeed, here electron-electron interactions within and among the channels at the edge of the quantum Hall bar can be treated exactly for integer edge state at filling factor $\nu=2$ in terms of the so called chiral Luttinger liquid formalism~\cite{Wen95, Delft98}, which predicts an interaction-induced fractionalization of an incoming time dependent voltage drive~\cite{Grenier13}.  

The present paper is organized as follows. In Section \ref{Model} we will describe the hybrid quantum thermal machine, discussing all the elements needed to properly characterize its functioning such as the edge-magnetoplasmon description of the quantum Hall edge channels and the photo-assisted tunneling through the quantum dot. Section \ref{Quantities} will present the charge, energy and heat currents which are essential building blocks needed to characterize the working regimes. Starting from the concept of entropy production rate, in Section \ref{exergy}, we will derive the general expression for the \textit{exergy}, the relevant figure of merit able to properly characterize the functioning of a hybrid device.  In Section \ref{Results} we will present results concerning the engine, the heat pump and a mixed working regime of the thermal machine using a train of Lorentzian voltage pulses as external time-dependent drive. We will also comment about the robustness of the results with respect electron-electron interaction among the quantum Hall channels and the possibility of enhancement of the exergy in a proper range of parameters. Section \ref{Conclusions} will be devoted to the conclusions. Finally, four Appendices will report some technical details of the calculations, a discussion about the choice of driving frequency and a brief account of the refrigerator regime.


\begin{figure}[H]
\centering
    \includegraphics[width=0.45\textwidth]{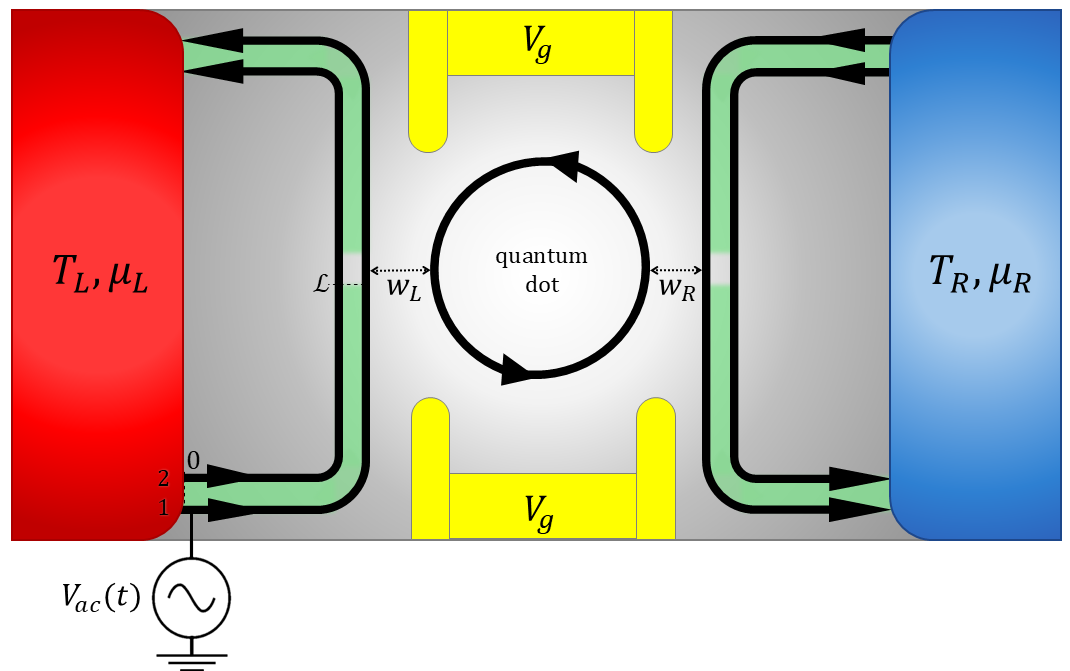}
    \caption{Scheme of the quantum Hall hybrid thermal machine. The terminals are quantum Hall states at filling factor $\nu=2$ kept at a given chemical potentials $\mu_{R}$, $\mu_{L}$ and temperatures $T_{R}$, $T_{L}$ respectively by means of two reservoirs (red and blue). The choice of colors has been done to visually show the assumption $T_{L}\geq T_{R}$. The left terminal $L$ is also driven out of equilibrium by applying an ac drive $V_{ac}(t)$ to its outer channel. Thanks to the action of gate voltages $V_{g}$ the two terminals are tunneling coupled with amplitudes $w_{R}$ and $w_{L}$ respectively to a Hall quantum dot at filling factor $\nu=1$. The numbers $1$ and $2$ denote the outer and inner channels respectively, and the green area represents the interaction region.}
    \label{Fig1}
\end{figure}

\section{Model}
\label{Model}

\subsection{Description of the set-up}
We consider a hybrid quantum thermal machine based on a Hall quantum dot, namely a system  with a discrete energy spectrum realized by a closed quantum Hall edge state, tunneling coupled with external right ($R$) and left ($L$) chiral edge states playing the role of terminals. They are coupled to reservoirs kept at different chemical potentials $\mu_{R}$ and $\mu_{L}$, and temperatures $T_{R}$ and $T_{L}$ respectively. In the following, we will assume $T_{L}\geq T_{R}$. Moreover, the outer channel of the $L$ terminal is also driven out-of-equilibrium by an additional periodic ac voltage (see Fig. \ref{Fig1}). We assume here that the gate voltages ($V_{g}$) are chosen in such a way that the quantum dot is composed by a unique closed channel (at filling factor $\nu=1$), while the terminals are at filling factor $\nu=2$, namely they are composed by two co-propropagating edge channels. In this configuration the electron-electron interaction among the edge states in the terminals need to be taken into account. In the following, we will analyze step by step all the relevant parts of this device in order to properly characterize its working regimes. 

\subsection{Interacting quantum Hall edge channels at $\nu=2$}

\begin{figure}[H]
\centering
    \includegraphics[width=0.45\textwidth]{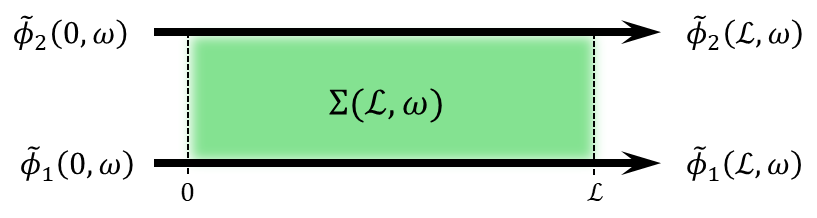}
    \caption{Scheme of a quantum Hall edge state at filling factor $\nu=2$. The interaction region of finite length $\mathcal{L}$ between the two channels
 is indicated by the green area. In the frequency domain, the incoming bosonic fields and the outgoing ones are denoted by $\tilde{\phi}_{1/2}(0,\omega)$ and $\tilde{\phi}_{1/2}(\mathcal{L},\omega)$ respectively and are connected by the edge-magnetoplasmon scattering matrix $\Sigma(\mathcal{L},\omega)$. }
    \label{Fig2}
\end{figure}

As usually done in the literature on the subject, we assume that each one of the two edge states at filling factor $\nu=2$, forming the terminals, is composed by 
one-dimensional ballistic channels interacting within a region of finite length $\mathcal{L}$ via a screened Coulomb repulsion~\cite{Sukhorukov07, Levkivskyi08, Degiovanni10, Grenier13,Wahl14, Ferraro14, Slobodeniuk16, Rebora20, Idrisov22}. This modelization has been crucial to explain various experimental observations~\cite{Altimiras10, Bocquillon13,Rodriguez20, Frigerio24}. To properly describe the system under this condition we need to consider the so called chiral Luttinger liquid picture~\cite{Wen95}, where the edge states are described in terms of bosonic modes chirally propagating along the edge, usually dubbed edge-magnetoplasmons~\cite{Degiovanni10, Wahl14, Sukhorukov16}. Within this picture the Hamiltonian density of each edge in the interacting region can be written as
\begin{equation}
    \mathcal{H}=\sum_{\alpha=1,2}\frac{\hbar v_i}{4\pi}\left[ \partial_x\phi_{\alpha}(x)\right]^2 +\frac{\hbar u}{2\pi}\left[ \partial_x\phi_{1}(x) \partial_x\phi_{2}(x)\right],
    \label{eq:H}
\end{equation}
where the index $\alpha=1,2$ labels the outer and inner channel respectively (see Fig.~\ref{Fig2}), while $\phi_{\alpha}$ are chiral bosonic fields satisfying the commutation relations
\begin{equation}
[\phi_\alpha(x),\phi_\beta(y)]=i \pi \,\mathrm{sign}(x-y) \delta_{\alpha, \beta}.
\end{equation} 
They are related to the particle density operator by
\begin{equation}
\rho_{\alpha}(x)=\frac{1}{2\pi}\partial_x \phi_{\alpha}(x).
\end{equation} 
The annihilation operator for an electron in the channel $\alpha=1,2$ can be written as a coherent state of the bosonic fields introduced above in such a way that 
\begin{equation}
    \psi_{\alpha}(x) = \frac{\mathcal{F}_{\alpha}}{\sqrt{2\pi a}}e^{-i\phi_{\alpha}(x)}
\end{equation}
with $a$ a short distance cut-off and $\mathcal{F}_{\alpha}$ playing the role of Klein factors~\cite{Wen95, Delft98, Crepieux04}. In Eq.~(\ref{eq:H}) $v_{1}, v_{2}$ are the bare propagation velocities of the two edge channels (satisfying $v_{1}\geq v_{2}$ under typical experimental conditions) and $u$ is the intensity of the density-density coupling between the edges. 

The Hamiltonian in Eq.~(\ref{eq:H}) can be diagonalized through a rotation in the bosonic space~\cite{Rebora20} of angle
\begin{equation}
\theta= \frac{1}{2}\arctan\left(\frac{2u}{v_1-v_2} \right).
\end{equation} 
This parameter encodes the interaction strength, indeed $\theta=0$ corresponds to the non-interacting case while $\theta=\pi/4$ represents the strong interacting limit~\cite{Levkivskyi08, Kovrizhin12, Braggio12}. The rotation leads to a charged and a neutral bosonic field defined respectively as
\begin{eqnarray}
\phi_c(x)&=&\phi_1(x) \cos\theta+\phi_2(x)\sin\theta\\
    \phi_n(x)&=&-\phi_1(x)\sin\theta+\phi_2(x)\cos\theta
\end{eqnarray}
with the full diagonalized Hamiltonian density assuming the form
\begin{equation}
\mathcal{H}=\frac{\hbar v_c}{4\pi}\left[\partial_x\phi_{c}(x)\right]^2+\frac{\hbar v_n}{4\pi}\left[\partial_x\phi_{n}(x)\right]^2\,
\end{equation}
with
\begin{equation}
    v_{c/n}=\left(\frac{v_1+v_2}{2}\right)\pm\frac{1}{\cos(2\theta)}\left(\frac{v_1-v_2}{2}\right).
\end{equation}

As schematically illustrated in Fig.~\ref{Fig2}, moving to the Fourier space (indicated in the following with the upper tilde notation) the dynamics of the edge channels can be solved within a scattering matrix formalism~\cite{Degiovanni10}, namely the component at frequency $\omega$ of the fields outgoing from an interacting region at finite length $\mathcal{L}$ is related to incoming ones by 
\begin{equation}
    \begin{pmatrix}
    \tilde{\phi}_{1}(\mathcal{L},\omega) \\
    \tilde{\phi}_{2}(\mathcal{L},\omega)
    \end{pmatrix}=\Sigma(\mathcal{L},\omega)\begin{pmatrix}
    \tilde{\phi}_{1}(0,\omega) \\
    \tilde{\phi}_{2}(0,\omega)
    \end{pmatrix}
    \label{eq:phi-out}
\end{equation}
with
\begin{equation}
\Sigma=
\begin{pmatrix}
\cos^2{\theta}\,e^{i\omega\tau_c}+\sin^2{\theta}\,e^{i\omega\tau_n} & \sin{\theta}\cos{\theta}(e^{i\omega\tau_c}-e^{i\omega\tau_n}) \\
\sin{\theta}\cos{\theta}(e^{i\omega\tau_c}-e^{i\omega\tau_n}) & \sin^2{\theta}\,e^{i\omega\tau_c}+\cos^2{\theta}\,e^{i\omega\tau_n}
 \end{pmatrix}.
 \label{eq:smatrix}
\end{equation}
Notice that, in the above expression we have introduced the times of flight $\tau_{c/n}=\mathcal{L}/v_{c/n}$.

We can now consider a capacitive coupling leading to the injection of a purely ac voltage $V_{ac}(t)$ with null average in a period only to the outer channels of the $L$ terminal (see Fig. \ref{Fig1}). This can be modeled by means of the additional contribution to the density Hamiltonian
\begin{equation}
\mathcal{H}_{U}=-e\rho_{1}(x)U(x,t),
\end{equation}
where $U(x,t)=\Theta(-x)V_{ac}(t)$ with $\Theta(-x)$ the Heaviside step function and $-e$ the electron charge~\cite{Rech17, Vannucci17}. 

This classical voltage creates a coherent state for the edge-magnetoplasmons along the edge channels~\cite{Safi99, Grenier13, Ferraro2018}. In the frequency domain the interacting region acts as a beam-splitter for this coherent state through the edge-magnetoplasmon scattering matrix $\Sigma(\mathcal{L},\omega)$ in exactly the same way as for the bosonic modes in absence of voltage. According to this, the voltages outgoing from the interacting region can be written as
\begin{equation}
    \begin{pmatrix}
    \tilde{V}_{1, out}(\omega) \\
    \tilde{V}_{2, out}(\omega)
    \end{pmatrix}=\Sigma(\mathcal{L},\omega)\begin{pmatrix}
    \tilde{V}_{ac}(\omega) \\
    0
    \end{pmatrix}.
    \label{eq:phi-out}
\end{equation}
Going back to the time domain this leads to
\begin{equation}
 \begin{split}
      V_{1,out}(t)=&\cos^2{\theta}\,V_{ac}(t-\tau_c)+\sin^2{\theta}\,V_{ac}(t-\tau_n)\\ 
      V_{2,out}(t)=&\sin{\theta}\cos{\theta}[V_{ac}(t-\tau_c)-V_{ac}(t-\tau_n)]\\ 
 \end{split}
 \label{eq:voltage}
\end{equation}
which is a manifestation of the so called fractionalization of the applied voltage~\cite{Grenier13, Ferraro14}. It is worth to note that, according to Eq.~(\ref{eq:voltage}), a dc voltage is not affected by the presence of interactions.  

Taking into account the above analysis, in the following we will consider the tunneling of electrons from the outgoing channels of the reservoirs to the quantum dot. This tunneling occurs at a distance $\mathcal{L}$ from the application of $V_{ac}(t)$ and is therefore characterized by the modified voltage $V_{1,out}(t)$. Moreover, it will be assumed to occur in absence of interactions due to the presence of additional screening of the Coulomb interaction at the level of the quantum point contacts illustrated in Fig.~\ref{Fig1}. Before doing this, however, it is useful to characterize the tunneling across the central quantum dot in absence of the applied time-dependent voltage.


\subsection{Tunneling through the quantum dot}

 As stated above, we consider a quantum dot made only by one closed channel ($\nu=1$). In this configuration, we also choose the size of the dot and consequently its level spacing in such a way that only one energy level, with given energy $E_{0}$, is relevant for our analysis. More specifically, due to the fact that the level spacing of the quantum dot scales as the inverse of its circumference, this situation is achieved if the size of the quantum dot is smaller than $1\,\,\mu m$, in this case the level spacing is greater than the other energy scales involved in the problem. Typically, for a dot circumference $\mathcal{C}\approx 100\,\,\mathrm{nm}$ and a propagation velocity of the edge modes of $v\approx 10^{5}\,\,\mathrm{m/s}$ one can estimate a level spacing 
 \begin{equation}
 \Delta E=\frac{\hbar v}{\mathcal{C}}\approx 10\,\, \mathrm{K}
 \end{equation}
 in temperature units. 

Considering a local electron tunneling occurring only at the level of the outer edge channels of the terminals and in absence of interactions, it is possible to introduce a new (fermionic) $2\times2$ scattering matrix $\mathbb{S}$ connecting the electrons incoming into the dot to the outgoing ones. It is unitary ($\mathbb{S}^\dagger \mathbb{S}=\mathbb{I}_{2\times 2}$) and assumes the general form 
\begin{equation}
    \mathbb{S}(E)=\begin{pmatrix}
        S_{LL}(E)&S_{LR}(E)\\
        S_{RL}(E)&S_{RR}(E)
    \end{pmatrix},
\end{equation}
where the notation $S_{jk}(E)$, with $j,k\in L, R$, keeps track of the part of the set-up from where the electron enters the dot ($k$) and to where it goes out ($j$). 

In order to proceed further, it is possible to write down an explicit form for the scattering matrix of the quantum dot~\cite{Jalabert1992, Alhassid2000}. Taking into account the fact that the dot has a unique relevant energy level and considering the tunneling amplitudes $w_{j}$ ($j=L, R$) for an electron to jump from the $j$-th terminal to the dot one can write
\begin{equation}
\label{S_matrix}
\begin{split}
\mathbb{S}(E)
&=\mathbb{I}_{2 \times 2} - \frac{2i\pi}{E-E_0+i\pi(|w_L|^2+|w_R|^2)}\begin{pmatrix}|w_L|^2&w_L^*w_R\\w_R^*w_L&|w_R|^2\end{pmatrix},
\end{split}
\end{equation} 
which satisfies the general properties discussed above. In particular, at a fixed energy, one has 
\begin{eqnarray}
    |S_{RL}(E)|^2&=&|S_{LR}(E)|^2=\frac{\Gamma_L\Gamma_R}{(E-E_0)^2+\frac{1}{4}(\Gamma_L+\Gamma_R)^2}\nonumber\\
    |S_{LL}(E)|^2&=&|S_{RR}(E)|^2=1-|S_{RL}(E)|^2,
    \label{constraints}
\end{eqnarray}
where we have introduced the short notation $\Gamma_{j}=2\pi|w_{j}|^2$. Here, $\Gamma_{j}/\hbar$ are the decay rates of the state of the dot associated to the emission of an electron into the $j$-th terminal. From the above expression, we notice that the transmission associated to the dot is a Breit-Wigner function peaked at $E=E_{0}$ and with width given by $\Gamma_L+\Gamma_R$. In the following, for sake of simplicity, we will assume a symmetric set-up with $\Gamma_L=\Gamma_R=\Gamma/2$, with $\Gamma$ considered as the unit for the energies, and we will consider $E_0=0$ as the reference of the energies. 

\subsubsection{Photo-assisted electron tunneling}
We want now to extend the previous discussion to include the effects of the periodic drive $V_{ac}(t)$ applied to the outer channel of the terminal $L$. It is worth to note that, in spite of the changes of the overall profile of the voltage due to interaction effects encoded in Eq.~(\ref{eq:voltage}), the periodicity of the drive affecting the electron before the tunneling is preserved, namely $V_{1,out}(t)=V_{1,out}(t+\mathcal{T})$ with $\mathcal{T}$ the period of the applied drive. This leads to the Fourier series

\begin{equation}
    e^{-i\frac{e}{\hbar}\int_0^t V_{1,\mathrm{out}}(t')dt'}=\sum_{l=-\infty}^{+\infty} \mathcal{P}_{l}(q)e^{-il\Omega t}
    \label{eq:def-ptilde}
\end{equation}
where $\Omega=2\pi/\mathcal{T}$ is the frequency of the drive and 
\begin{equation}
q=\frac{-e V_{0}}{\hbar \Omega}
\end{equation}
with $V_{0}$ the amplitude of the incoming ac drive (see Sec. \ref{Results} for more details about the ac drive). In the above expression the coefficients $\mathcal{P}_{l}$ are the probability amplitudes for an electron to absorb (emit) $l$ photons for $l>0$ ($l<0$) due to the drive~\cite{Dubois13}. They can be written as~\cite{Rebora20} 

\begin{equation}
    \begin{split}
    \mathcal{P}_{l}(q)&=\sum_{k=-\infty}^{+\infty}p_{l-k}(q\cos^2\theta)p_n(q\sin^2\theta)\\
    &\quad\times e^{i\Omega\tau_c(l-k)}e^{i\Omega\tau_n(k)},
    \end{split}
    \label{eq:coeff2}
\end{equation}
where the coefficients $p_{l}$ are defined as 
\begin{equation}
p_{l}(q)=\frac{\Omega}{2 \pi} \int^{\frac{2 \pi}{\Omega}}_{0} dt e^{-i \frac{e}{\hbar} \int^{t}_{0} dt' V_{ac}(t')}e^{i l \Omega t}.
\label{p_l}
\end{equation}

Within this photo-assisted tunneling picture it is possible to extend the previous analysis by introducing the so called Floquet scattering matrix for the dot in presence of a driven lead, given by~\cite{Moskalets08, Moskalets14, Moskalets_Book} 
\begin{eqnarray}\label{S^F}
    \mathbb{S}^{(F)}(E_{n}, E)&=&\begin{pmatrix}
        S^{(F)}_{LL}(E_{n}, E)&S^{(F)}_{LR}(E_{n}, E)\\
        S^{(F)}_{RL}(E_{n}, E)&S^{(F)}_{RR}(E_{n}, E)
    \end{pmatrix}\nonumber\\
    &=&
    \begin{pmatrix}
        \mathcal{P}_{n}(q)S_{LL}(E_{n}) & \delta_{n, 0} S_{LR}(E)\\
        \mathcal{P}_{n}(q) S_{RL}(E_{n}) & \delta_{n, 0}S_{RR}(E) 
    \end{pmatrix},
\end{eqnarray}
with $E_{n}=E+n \hbar \Omega$. Notice that in the above expression the effect of interaction is encoded in the coefficients $ \mathcal{P}_{n}(q)$. For further convenience, it is also useful to introduce an equivalent expression properly derived from the one above by replacing $E\rightarrow E-n\hbar \Omega $ and  $n\rightarrow -n$, namely 

\begin{eqnarray}\label{S^F}
    \mathbb{S}^{(F)}(E, E_{n})&=&\begin{pmatrix}
        \mathcal{P}_{-n}(q)S_{LL}(E) & \delta_{n, 0} S_{LR}(E_n)\\
        \mathcal{P}_{-n}(q) S_{RL}(E) & \delta_{n, 0}S_{RR}(E_n) 
    \end{pmatrix}.
\end{eqnarray}

In the following, we will report the expressions for the relevant physical quantities (currents) needed to characterize the working regimes of such hybrid thermal machine within the Floquet scattering matrix formalism. 


\section{CURRENTS}
\label{Quantities}
\subsection{Charge currents}
The dc (averaged with respect to the ac drive period $\mathcal{T}$) contribution to the charge current flowing between the $j$-th ($j=L, R$) lead and the dot is given by~\cite{Moskalets08} 

\begin{equation}
I^{e}_{j}=-\left(\frac{e}{2 \pi \hbar}\right) \int^{+\infty}_{-\infty} d E \left[f^{out}_{j}(E)-f_{j}(E) \right] 
\label{charge_current}
\end{equation}
with $-e$ ($e>0$) the electron charge, 
\begin{equation}
f_{j}(E)=\frac{1}{1+ e^{\frac{E-\mu_{j}}{k_{B}T_{j}}}}
\end{equation}
the equilibrium Fermi distribution associated to the $j$-th reservoir and 
\begin{equation}
    f_{j}^{out}(E)=\sum_{n=-\infty}^{+\infty}\sum_{k=L,R}|S^{(F)}_{j k}(E,E_n)|^2 f_{k}(E_n)
    \label{f_j_out}
\end{equation}
the out-of equilibrium distribution created by the Floquet scattering. 

Notice that with the considered definitions a positive current is outgoing from the dot and a negative one otherwise. 

By properly exploiting the unitary of the Floquet scattering matrix 
one can write the charge currents as 

\begin{eqnarray}
I_L^e&=&-\left(\frac{e}{2 \pi \hbar}\right)\int_{-\infty}^{+\infty} dE\, \left[\tau_{LR}(E)f_R(E)-\tau_{RL}(E)f_L(E)\right],\nonumber\\
\label{IeL} \\
I_R^e &=& 
-\left(\frac{e}{2 \pi \hbar}\right)\int_{-\infty}^{+\infty} dE \left[\tau_{RL}(E)f_L(E)-\tau_{LR}(E)f_R(E)\right]\nonumber\\
\label{IeR}
\end{eqnarray}
where we have introduced the photo-assisted transmission amplitudes
\begin{eqnarray}
    \tau_{RL}(E)&\equiv& \sum_n |S_{RL}^{(F)}(E_n,E)|^2 \label{TRL}\\ 
    \tau_{LR}(E)&\equiv& \sum_n |S_{LR}^{(F)}(E_n,E)|^2.
    \label{TLR}
\end{eqnarray}

It is worth noting that the charge currents satisfy 
\begin{equation}
I_L^e+I_R^e=0.
\label{charge_conservation}
\end{equation} 

\subsection{Energy currents}

Proceeding as above, it is also possible to define the dc contribution to the energy currents as 

\begin{equation}
I^{u}_{j}=\int^{+\infty}_{-\infty} \frac{d E}{2 \pi \hbar}E\left[f^{out}_{j}(E)-f_{j}(E) \right].
\label{I^u}
\end{equation}
Notice that the above expression is obtained from Eq.~(\ref{charge_current}) by formally replacing the electric charge $-e$ carried by each electron with its energy $E$. By further manipulating the above expressions (see details of the calculation in Appendix \ref{App1}) one obtains

\begin{eqnarray}
I_L^u&=&\int_{-\infty}^{+\infty} \frac{dE}{2 \pi \hbar}E[\tau_{LR}(E)f_R(E)-\tau_{RL}(E)f_L(E)]+\label{IuL}\nonumber\\
&+&\frac{\Omega}{2\pi}\sum_{k=L,R}\int_{-\infty}^{+\infty} dE\langle n(E)\rangle_{Lk}f_k(E)+P_{in}, \label{IuL}\\
I_R^u&=&\int_{-\infty}^{+\infty} \frac{dE}{ 2 \pi \hbar} E[\tau_{RL}(E)f_L(E)-\tau_{LR}(E)f_R(E)]+\nonumber\\
&+&\frac{\Omega}{2\pi}\sum_{k=L,R}\int_{-\infty}^{+\infty} dE\langle n(E)\rangle_{Rk}f_k(E).\label{IuR}
\end{eqnarray}
Here,
\begin{equation}\label{n_medio}
    \langle n(E)\rangle_{jk}=\sum_n n|S_{jk}^{(F)}(E_n,E)|^2
\end{equation}
is the average number of photons absorbed or emitted during the scattering process associated to the $j$-th lead and 
\begin{equation}
P_{in}=\frac{e^{2}}{4 \pi \hbar} \frac{1}{\mathcal{T}} \int^{+\frac{\mathcal{T}}{2}}_{-\frac{\mathcal{T}}{2}} dt V^{2}_{1, out}(t)\geq 0
\label{P_in}
 \end{equation}
 the dc power due to the time-dependent drive. 
 
 Also in this case the currents are positive if outgoing from the dot and negative otherwise.  Taking into account the fact that for a purely ac drive\cite{Dubois13, Vannucci17, Ryu22}
 \begin{equation}
 \sum_{n}n |\mathcal{P}_{n}|^{2}=\left(\frac{e}{\hbar \Omega}\right)\frac{1}{\mathcal{T}}\int^{+\frac{\mathcal{T}}{2}}_{-\frac{\mathcal{T}}{2}} V_{1, out}(t) dt =0
 \end{equation}
in this case one has that 
\begin{equation}
I_L^u+I_R^u=P_{in}.
\label{energy_conservation}
\end{equation}
 
\subsection{Heat currents}

The heat currents are written in terms of the energy and charge ones as 

\begin{equation}
I^{h}_{j}=I^{u}_{j}+\frac{\mu_{j}}{e}I^{e}_{j}
\label{heat_current}
\end{equation}

and consequently they are given by~\cite{Moskalets14, Battista14} 
\begin{eqnarray}
I_L^h&=&\int_{-\infty}^{+\infty} \frac{dE}{2 \pi \hbar}(E-\mu_L)[\tau_{LR}(E)f_R(E)-\tau_{RL}(E)f_L(E)]+\nonumber\\
&+&\frac{\Omega}{2\pi}\sum_{k=L,R}\int_{-\infty}^{+\infty} dE\langle n(E)\rangle_{Lk}f_k(E)+P_{in}, 
\label{IhL} \\
I_R^h&=&\int_{-\infty}^{+\infty} \frac{dE}{2 \pi \hbar} (E-\mu_R)[\tau_{RL}(E)f_L(E)-\tau_{LR}(E)f_R(E)]+\nonumber\\
&+&\frac{\Omega}{2\pi}\sum_{k=L,R}\int_{-\infty}^{+\infty} dE\langle n(E)\rangle_{Rk}f_k(E).\label{IhR}
\end{eqnarray}

Together they satisfy 
\begin{eqnarray}
I_L^h+I_R^h&=& P_{in}+\frac{\mu_{R}-\mu_{L}}{e}I^{e}_{R}\\
&=&P_{in}-P_{e}.
\end{eqnarray}
Here, the second term  
\begin{equation}
P_{e}=\frac{\mu_{L}-\mu_{R}}{e}I^{e}_{R}
\end{equation}
represents the electric power. Notice that for $P_{e}>0$ we create a useful work due to the fact that we induce a particle current flowing agains the direction imposed by the chemical potential.

In terms of the quantities discussed in this Section it is possible to derive a convenient figure of merit able to characterize the performances of our device as a hybrid thermal machine. 


\section{Entropy production rate and exergy}
\label{exergy}

We recall now the conventional definition for the entropy production rate~\cite{Benenti17}

\begin{equation}
\dot{\mathcal{S}}=\frac{I^{h}_{L}}{T_{L}}+\frac{I^{h}_{R}}{T_{R}}.
\end{equation}
Taking into account the current conservation constraints in Eqs.~(\ref{charge_conservation}) and (\ref{energy_conservation}), together with the definition of heat current in Eq.~(\ref{heat_current}), the above expression can be conveniently rewritten as 

\begin{equation}
\dot{\mathcal{S}}=I^{h}_{L}\left(\frac{1}{T_{L}} -\frac{1}{T_{R}}\right)+\frac{P_{in}-P_{e}}{T_{R}}.
\end{equation}

It is now useful to separate the positive ($\dot{\mathcal{S}}_{i}^{(+)}$) and the negative ($\dot{\mathcal{S}}_{i}^{(-)}$) contributions to the entropy production rate. According to this, the efficiency of a hybrid thermal machine can be determined in terms of the so called \emph{exergy}~\cite{Manzano20, Lu23, Lopez23} defined as the ratio between the negative contributions changed in sign, and the positive contributions, namely  
\begin{equation}
\Phi=- \frac{\sum_{i} \dot{\mathcal{S}}^{(-)}_{i}}{\sum_{i} \dot{\mathcal{S}}^{(+)}_{i}}.
\end{equation}
This figure of merit is relevant to quantify the performance of a multi-terminal thermal machine in a unified fashion taking also into account the possibility of multitasking configurations. It is normalized in such a way that $0\leq \Phi\leq 1$ and, for a two-terminal set-up under stationary conditions, reduces to the conventional thermodynamic efficiency in units of the Carnot efficiency~\cite{Cavaliere23}.

In order to properly characterize the relevant working regimes of the device under investigation (see Sec.~\ref{Results} below) we can further specify the relation as  

\begin{equation}
\Phi=\frac{I^{h}_{L}\Theta(I^{h}_{L})\left(1-\frac{T_{R}}{T_{L}}\right)+P_{e}\Theta(P_{e})}{P_{in}-I^{h}_{L}\Theta(-I^{h}_{L})\left(1-\frac{T_{R}}{T_{L}}\right)-P_{e}\Theta(-P_{e})},
\label{exergy_hybrid}
\end{equation}
with $\Theta(x)$ the Heaviside Theta function and where we have taken into account the fact that, while $P_{in}$ is always positive, $I^{h}_{L}$ and $P_{e}$ can change sign depending on the considered working point. 

In the following, we will discuss the results concerning relevant functioning regimes of the proposed device such as engine, heat pump and mixed regime: moreover, we will comment on their robustness with respect to electron-electron interaction. 


\section{Results}
\label{Results}

\subsection{Time-dependent drive}

We now have all the ingredients needed to investigate the working regimes of the proposed hybrid thermal machine. To fix the ideas, in the following we will consider as applied periodic drive $V_{ac}(t)$ a train of Lorentzian pulses, of width $\mathcal{W}$ and period $\mathcal{T}$, of the form
\begin{equation}\label{Vac}
V_{ac}(t)=V_0\left[\frac{1}{\pi}\sum_{k=-\infty}^{+\infty}\frac{\delta}{\delta^2+\left(\frac{t}{\mathcal{T}}-k\right)^2}-1\right],
\end{equation}
where $\delta=\mathcal{W}/\mathcal{T}$ and such that $-e V_{0}= q \hbar \Omega$, with $q$ a parameter that will be specified in the following. Note that this potential has zero average over one period, as expected for a purely ac drive. 

As theoretically predicted by Levitov and coworkers~\cite{Levitov96,Levitov97,Levitov06}and experimentally demonstrated by D. C. Glattli and his group~\cite{Glattli2013,Glattli2016,Glattli2018, Aluffi23} this kind of drive, in absence of interaction and by imposing the quantization condition $q\in\mathbb{N}$,
leads to the injection of purely electronic excitations, usually dubbed Levitons, without introducing in the system any electron-hole pair. In the following, for simplicity, we will set $q=1$ even if qualitatively analogous results can be obtained for different values of $q$. Moreover, the width of the pulses and the period will be chosen within an experimentally achievable range~\cite{Glattli17}.

For a periodic potential of the form in Eq.~\eqref{Vac} the coefficients $p_l(q)$ defined in Eq.~(\ref{p_l}) are given by~\cite{Dubois13,Grenier13,Rech17,Ferraro18}
\begin{equation}
p_l(q)=q\sum_{s=0}^{+\infty}\frac{(-1)^s {\boldsymbol{\Gamma}}(q+l+s)e^{-2\pi\delta(2s+l)}}{{\boldsymbol{\Gamma}}(q+1-s){\boldsymbol{\Gamma}}(1+s){\boldsymbol{\Gamma}}(1+l+s)},
\label{eq:coefflor}
\end{equation}
with ${\boldsymbol{\Gamma}}(...)$ the Euler's Gamma function. Thus, the photo-assisted amplitudes $\mathcal{P}_{l}(q)$ are completely specified by Eq.~\eqref{eq:coeff2} (fixing $q=1$ as stated above).

\subsection{Engine regime}
\begin{figure*}
    \centering
    \subfloat[]{
        \includegraphics[width=0.31\textwidth]{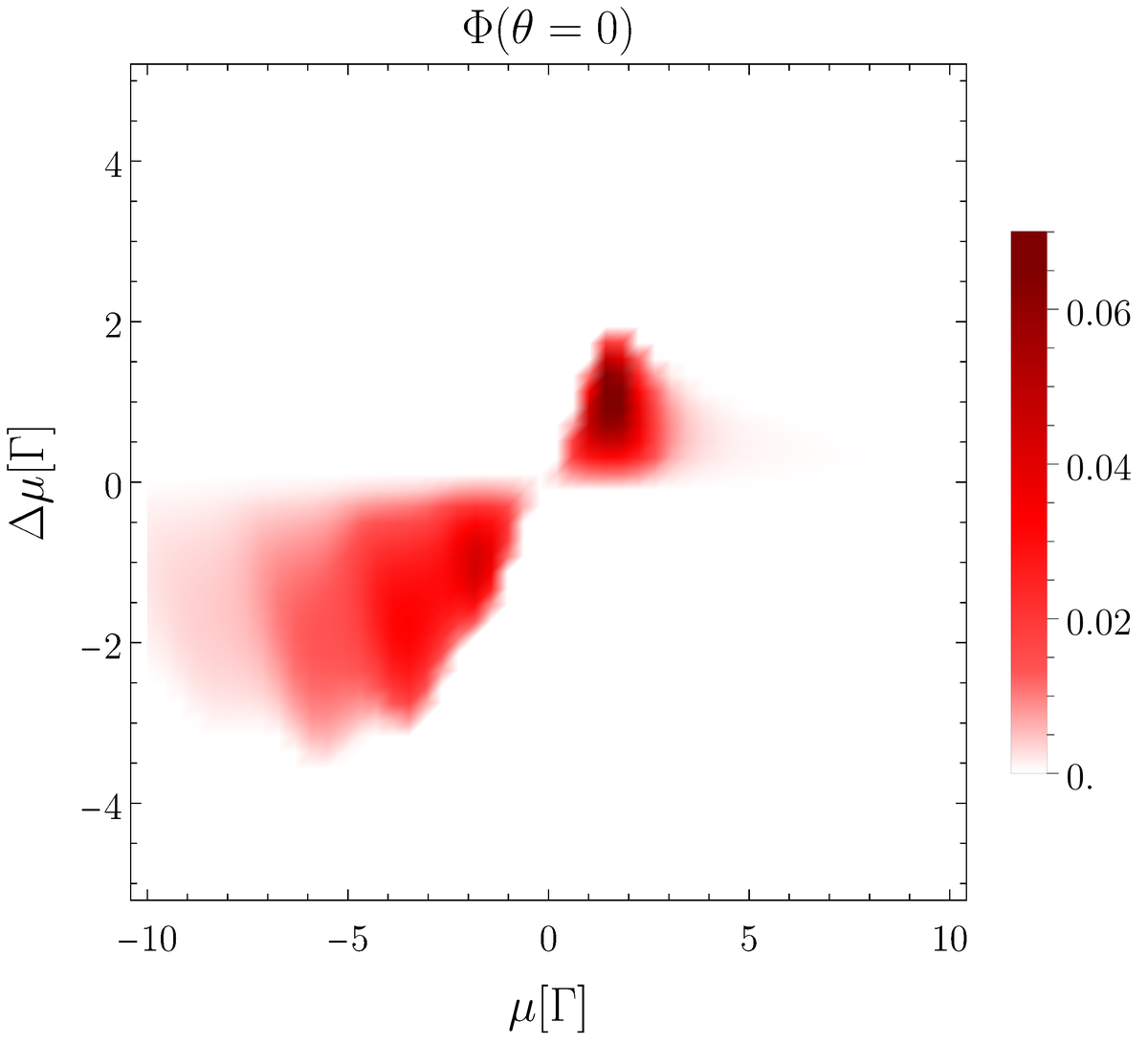}
        }
    \hfill
    \subfloat[]{
        \includegraphics[width=0.31\textwidth]{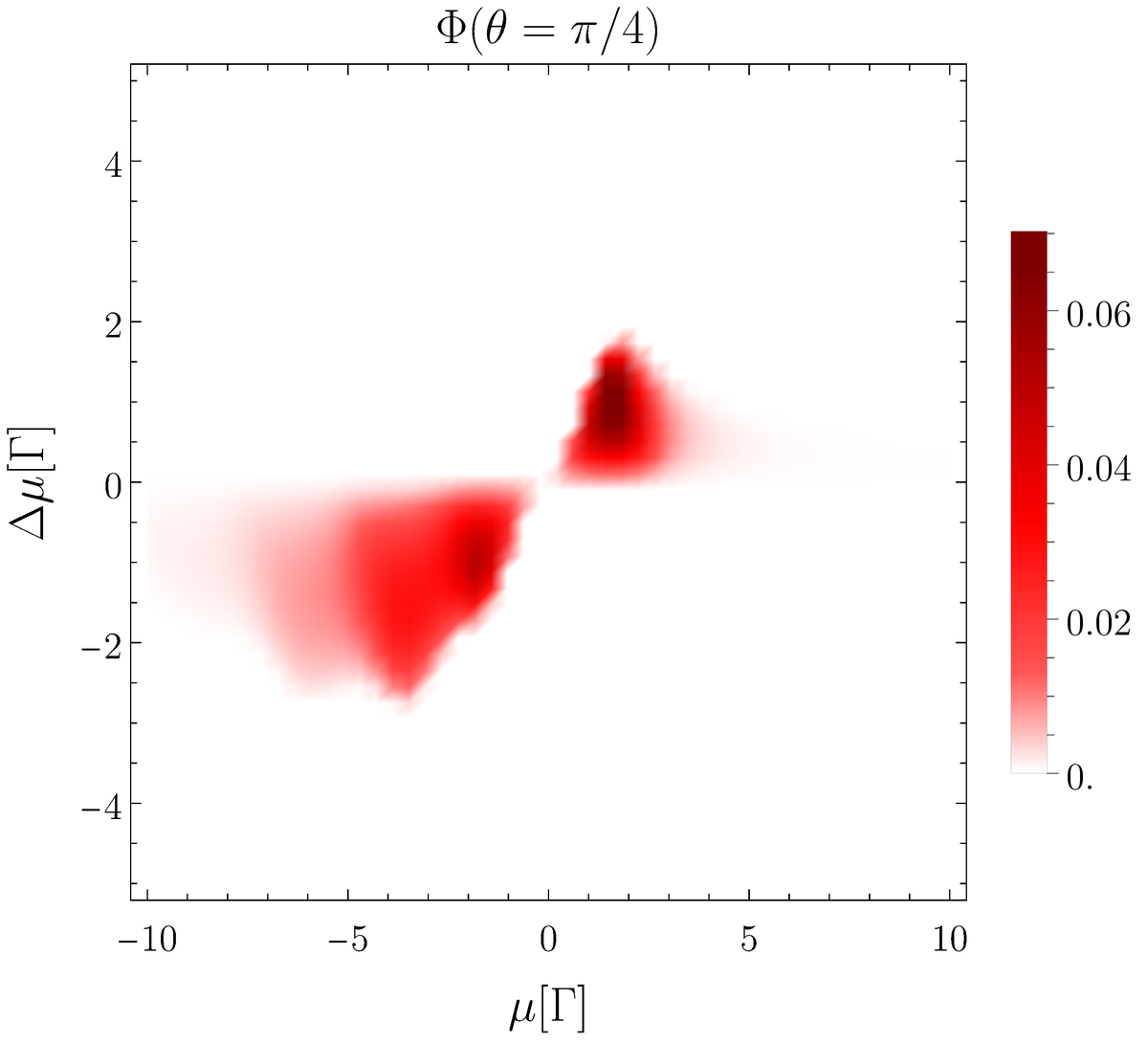}
        }
    \hfill
    \subfloat[]{
        \includegraphics[width=0.31\textwidth]{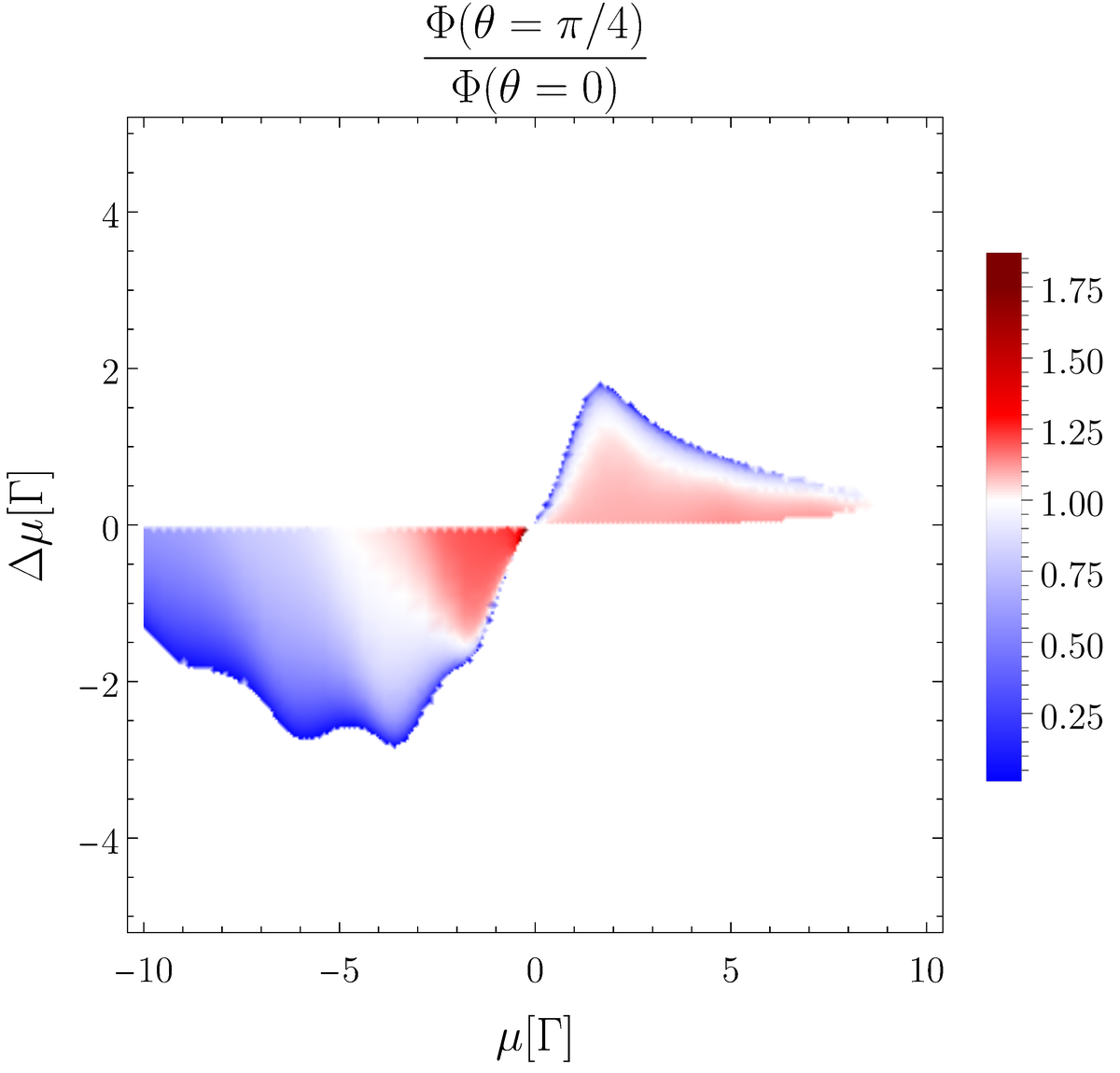}
        }
    \caption{Density plots of the exergy $\Phi$ in the engine regime ($P_{e}>0$)  as a function of $\mu$ and $\Delta\mu$ (in units of $\Gamma$), for the non-interacting case ($\theta=0$), in panel (a), and with interactions ($\theta=\pi/4$), in panel (b). Panel (c) shows the ratio between the exergy in the interacting and non-interacting case, $\Phi(\pi/4)/\Phi(0)$.
    The other parameters are: $T_R=T_L=0.01\,\Gamma/k_B$, $\Omega=2.6 \,\Gamma/\hbar$, $\tau_c=7.5\times 10^{-12}\, s$, $\tau_n=1.5\times 10^{-11}\, s$, $q=1$ and $\delta=0.09$.}
    \label{fig:Phi_engine}
\end{figure*}

In this case we fix $T_{L}=T_{R}$ and focus on the condition $P_{e}>0$. Here, the exergy in Eq.~(\ref{exergy_hybrid}) reduces to 
\begin{equation}
\Phi=\frac{P_{e}}{P_{in}},
\end{equation}
a typical work-to-work efficiency. 
In Figure \ref{fig:Phi_engine} we show the behavior of this quantity as a function of the average chemical potential
\begin{equation}
\mu=\frac{\mu_{L}+\mu_{R}}{2}
\end{equation}
and chemical potential difference
\begin{equation}
\Delta \mu= \mu_{L}-\mu_{R}.
\end{equation}
Notice that, for completeness, the behavior of $P_{e}$ in the regions where it is positive is reported in Appendix~\ref{App2}. We consider both the non-interacting ($\theta=0$) and the strongly interacting case ($\theta=\pi/4$). First of all, we observe that in the non-interacting case ($\theta=0$) there are some regions of the density plot where the presented device can be seen as a work-to-work engine, namely it is able to convert the power injected from the ac drive into positive electric power (see Fig. \ref{fig:Phi_engine} (a)). For the value of the drive frequency considered, one has $\Phi\approx 7\%$ (for lower and higher frequencies the efficiency is smaller, as shown in Appendix~\ref{App3}). The behavior is qualitatively the same for strong electron-electron interaction between the edge channels ($\theta=\pi/4$) as a signature of the robustness of the phenomenology (see Fig. \ref{fig:Phi_engine} (b)). 
Moreover, it is also possible to identify regions where the interaction leads to an enhancement of the exergy at parity of all the other parameters (see red regions in Fig. \ref{fig:Phi_engine} (c)).

\begin{figure*}
    \centering
    \subfloat[]{
        \includegraphics[width=0.31\textwidth]{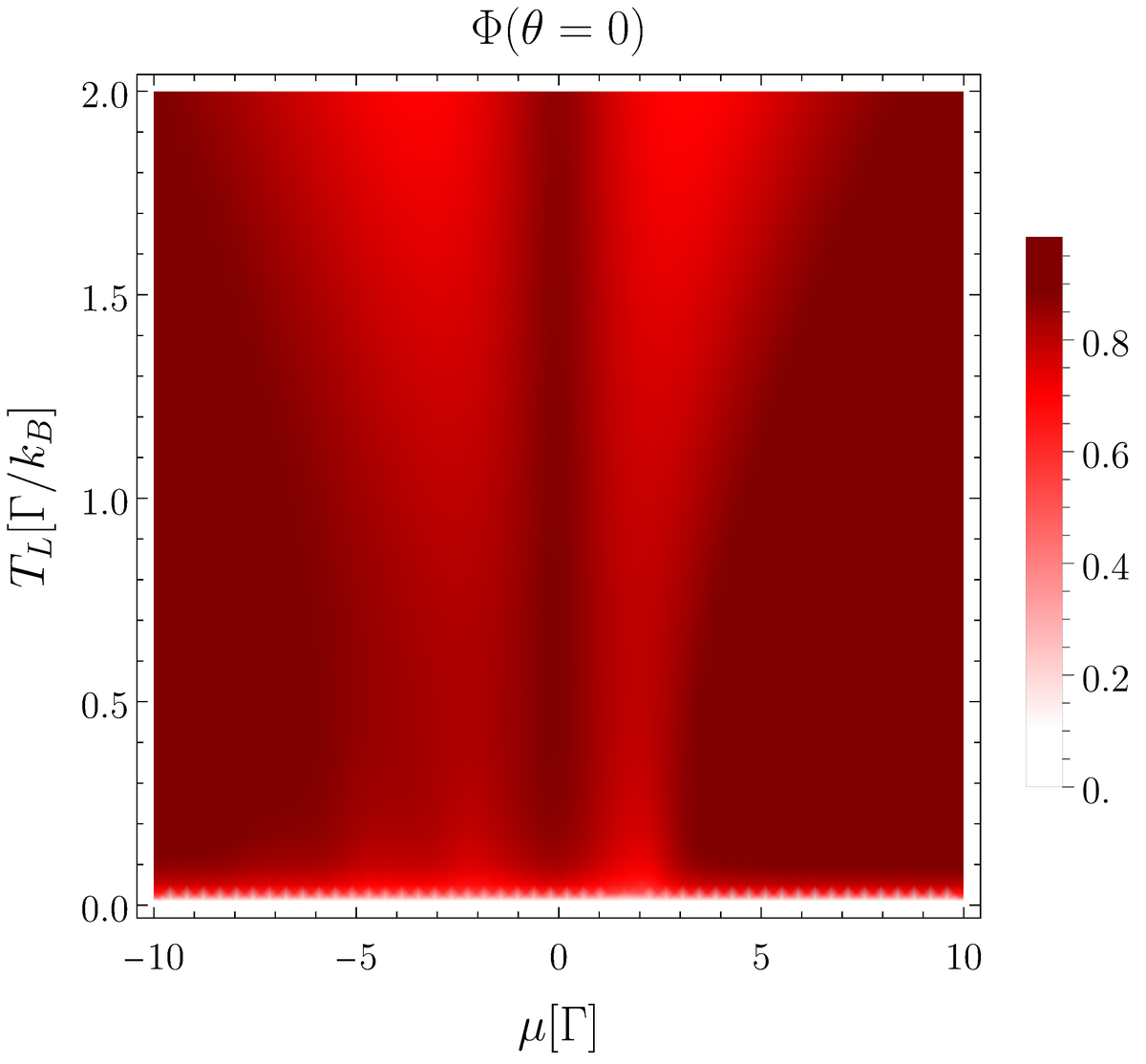}
        }
    \hfill
    \subfloat[]{
        \includegraphics[width=0.31\textwidth]{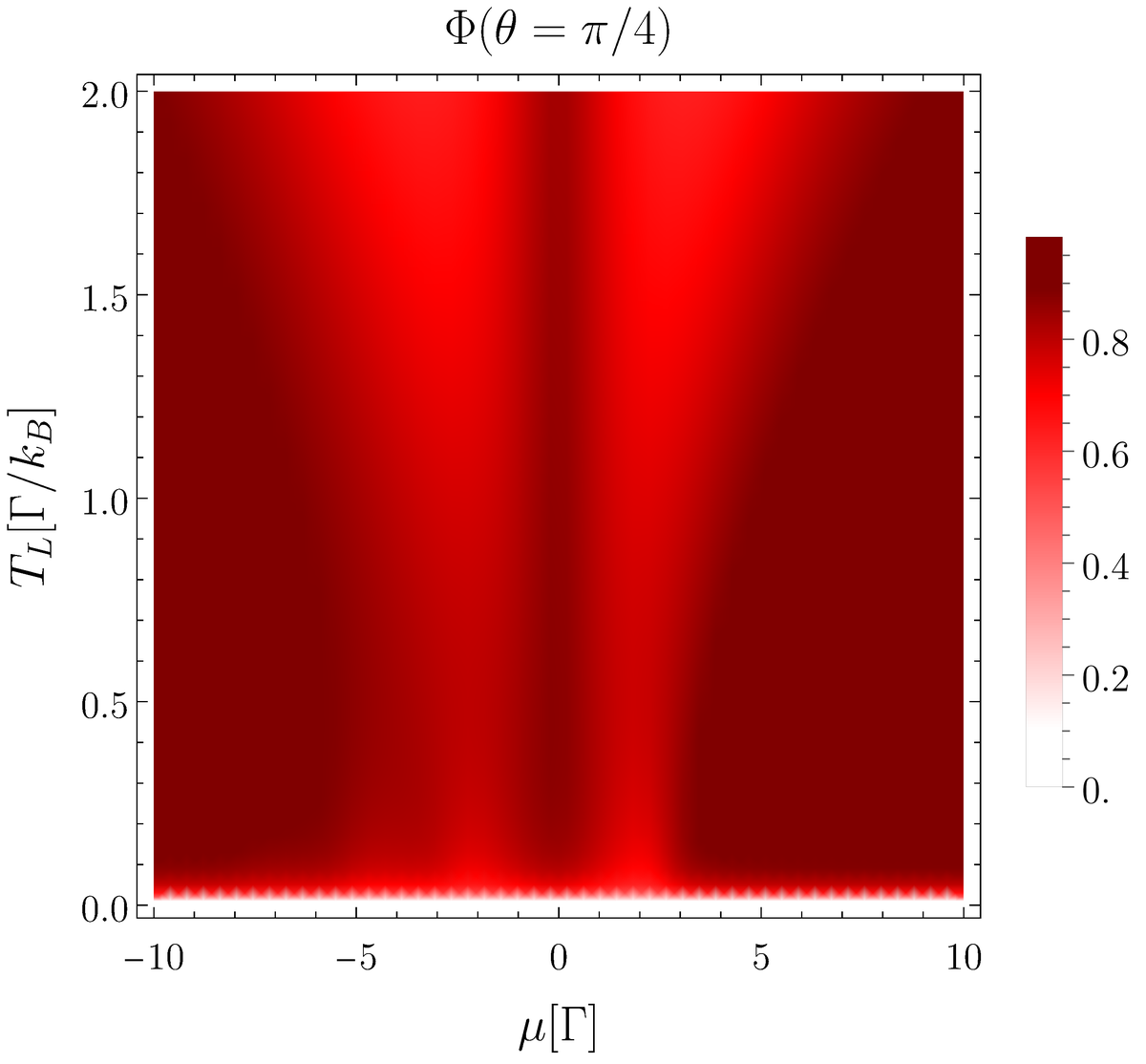}
        }
    \hfill
    \subfloat[]{
        \includegraphics[width=0.31\textwidth]{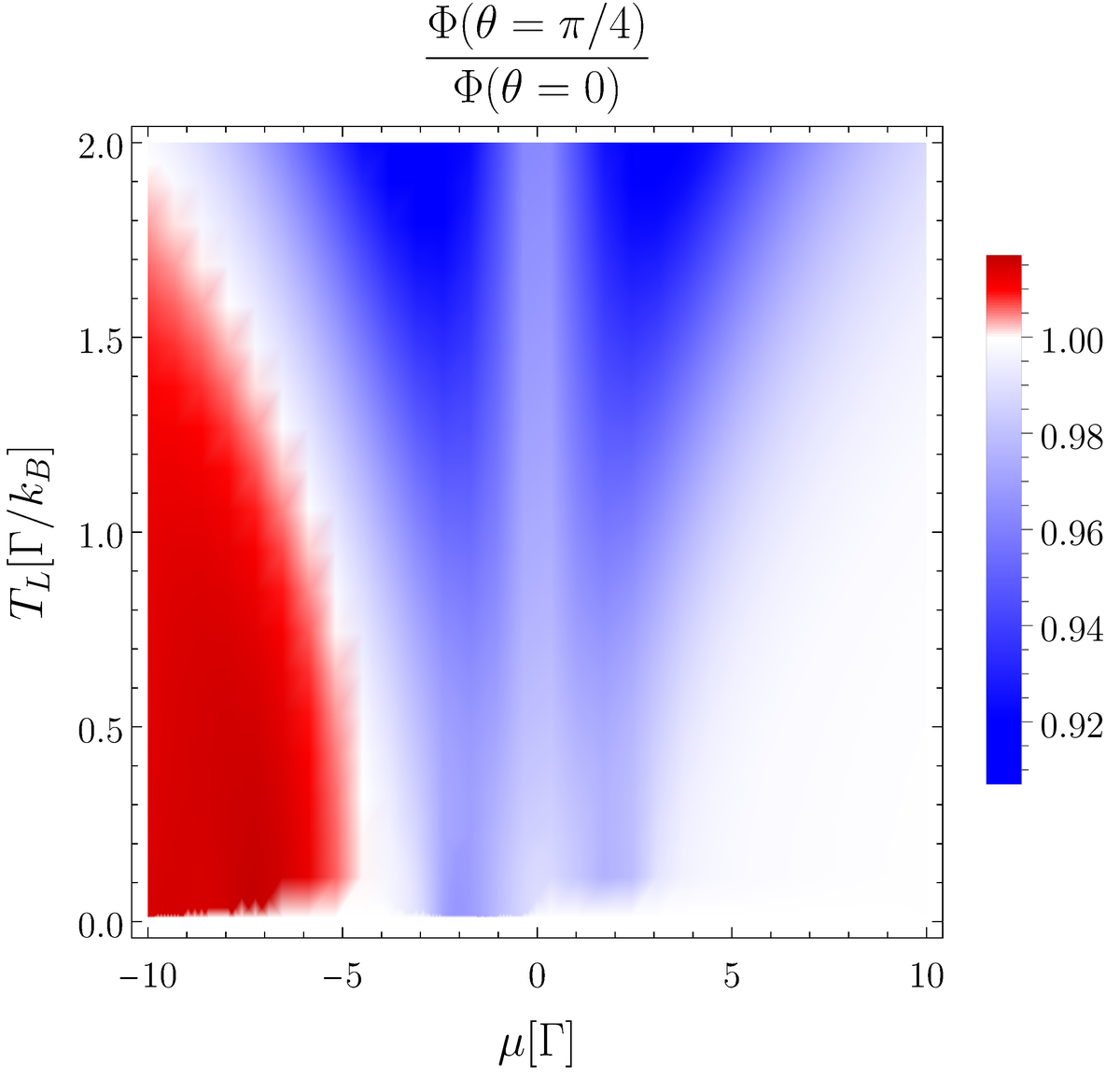}
        }
    \caption{Density plots of the exergy $\Phi$ in the heat pump regime ($I^{h}_{L}>0$) as a function of $\mu$ (in units of $\Gamma$) and $T_L$ (in units of $\Gamma/k_B$) for the non-interacting case ($\theta=0$), in panel (a), and with interactions ($\theta=\pi/4$), in panel (b). Panel (c) shows the ratio between the exergy in the interacting and non-interacting case, $\Phi(\pi/4)/\Phi(0)$.
    The other parameters are: $T_R=0.01\,\Gamma/k_B$, $\Delta\mu=\Gamma$, $\Omega=2.6 \,\Gamma/\hbar$, $\tau_c=7.5\times 10^{-12}\, s$, $\tau_n=1.5\times 10^{-11}\, s$, $q=1$ and $\delta=0.09$}
    \label{fig:Phi_pump}
\end{figure*}

\subsection{Heat pump regime}

Here, we consider the heat pump regime with $\mu_{L}=\mu_{R}$, focusing on $I^{h}_{L}>0$. This leads to the simplified expression for the exergy in Eq.~(\ref{exergy_hybrid}), since $P_{e}=0$,

\begin{equation}
\Phi=\frac{I^{h}_{L}}{P_{in}}\left(1-\frac{T_{R}}{T_{L}}\right).
\end{equation}
This quantity is shown in Figure \ref{fig:Phi_pump} as a function of $\mu$ and $T_{L}$ at fixed $T_{R}$. Also in this case, for completeness, the density plot of $I^{h}_{L}$ alone is reported in Appendix~\ref{App2}. Both in the non-interacting case (see Fig. \ref{fig:Phi_pump} (a)) and in the strong interacting case (see Fig. \ref{fig:Phi_pump} (b)) one observes the heat pump regime for all the considered values of the density plot with $\Phi\approx 90\%$. This suggests that, in our device, the heat pump behavior is the dominant working regime. This point will be better clarified in the following. The two panels are qualitatively very similar and, as before, it is possible to identify a region in the parameters space where the interaction leads to a slight (few percent) enhancement of the efficiency (see red regions in Fig. \ref{fig:Phi_pump} (c)). Due to the fact that a similar robustness with respect to electron-electron interactions between the edge channels is ubiquitous for all the working regimes discussed in this paper from now on we will focus only on the non-interacting case in order to keep the discussion more compact.

\subsection{Hybrid regime}

Combining the conditions $T_{L}> T_{R}$ and $\mu_{L}\neq \mu_{R}$, the engine and heat pump behavior can coexist leading to an interesting hybrid thermal machine. To describe it, we need to consider the complete expression for the exergy in Eq.~(\ref{exergy_hybrid}) shown as a function of $\mu$ and $\Delta \mu$ in Figure \ref{fig:ibrido_Dmu} (a). Here, the exergy is very close to saturation ($\Phi\approx 95\%$). The relative weight of the engine contribution ($\Phi_{eng}$) with respect to the heat pump contribution ($\Phi_{pump}$) in this hybrid regime can be seen in Figure \ref{fig:ibrido_Dmu} (b). It approaches a maximum of $\Phi_{eng}/\Phi_{pump}\approx 9\%$ justifying the dominant nature of the heat pump behavior 
suggested above. It is also instructive to describe this mixed regime as a function of $\mu$ and $T_{L}$ at a fixed $\Delta \mu$ and $T_{R}$. Here, the coexistence of the two behaviors occurs at $\mu>0$ (see Figure \ref{fig:ibrido_DT} a) and the relative weight of the engine contribution can reach $\Phi_{eng}/\Phi_{pump}\approx 17\%$ (see Figure \ref{fig:ibrido_DT} b). As a final comment, it is worth to note that the proposed device is also characterized by a refrigerator regime which will be discussed in Appendix~\ref{App4}.

\begin{widetext}
\begin{figure*}
    \centering
    \subfloat[]{
        \includegraphics[width=0.38\textwidth]{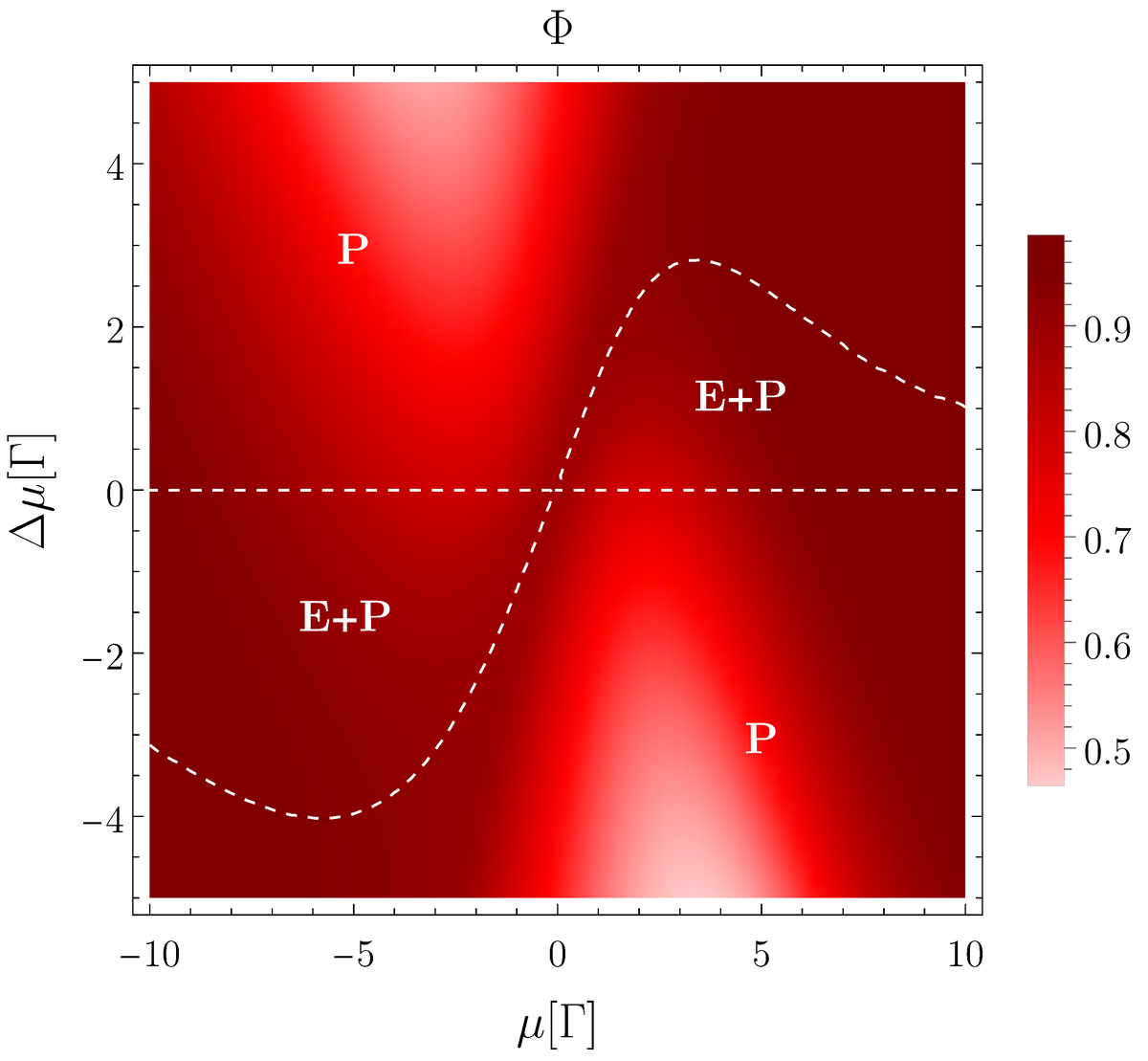}
        }
        \hspace{0.5cm}
    \subfloat[]{
        \includegraphics[width=0.38\textwidth]{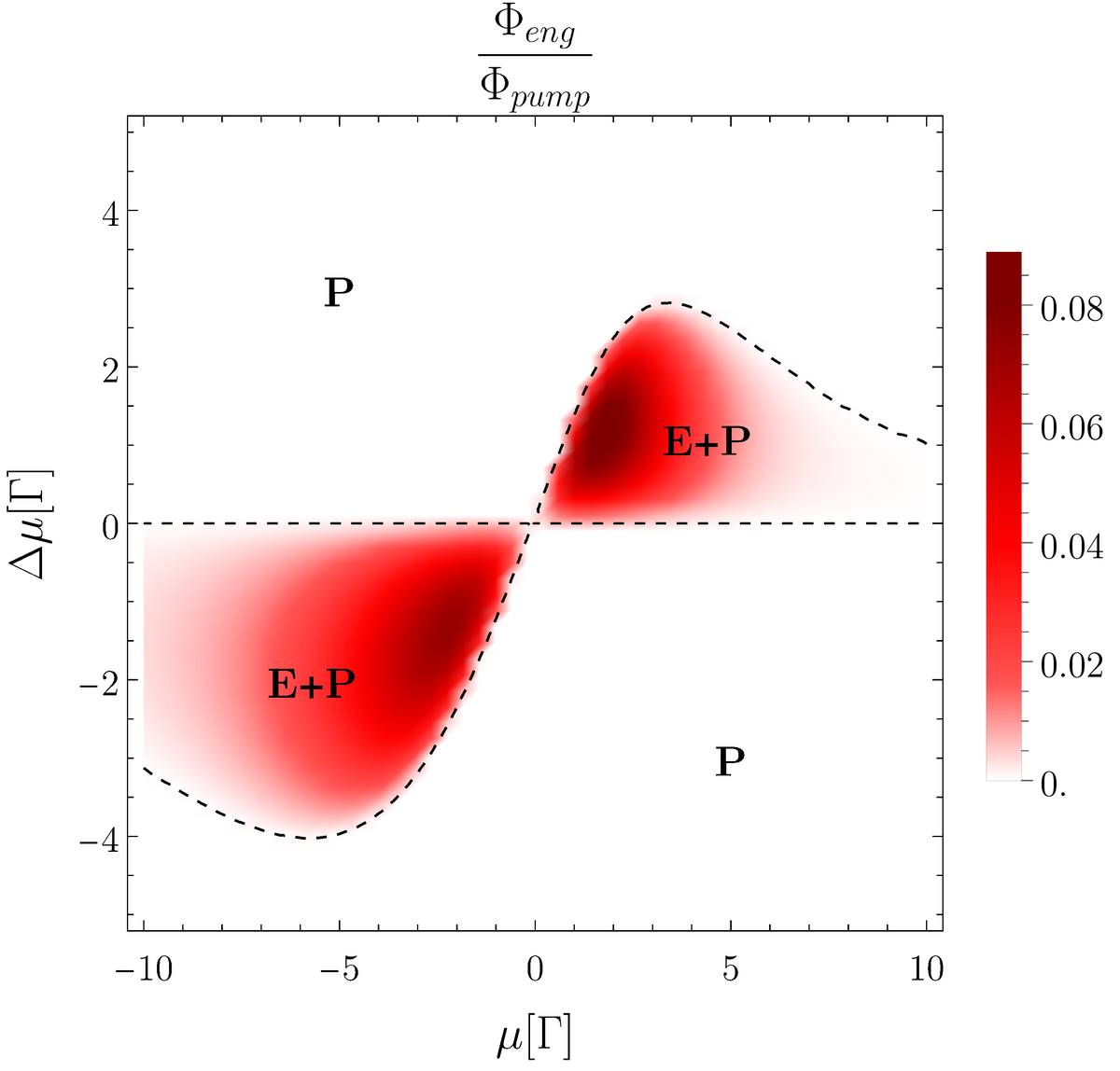}
        }
    \caption{(a) Density plot of exergy $\Phi$ as a function of $\mu$ and $\Delta \mu$ (in units of $\Gamma$). (b) Density plot of the ratio between the engine contribution and the heat pump contribution to the exergy, as a function of $\mu$ and $\Delta \mu$ (in units of $\Gamma$). In the regions marked with P the system acts as a heat pump, while the regions of coexistence of the two behavior are marked with E+P. The other parameters are: $T_R=0.01\,\Gamma/k_B$, $T_L=1.01\, \Gamma/k_B$, $\Omega=2.6 \,\Gamma/\hbar$, $\tau_c=7.5\times 10^{-12}\, s$, $\tau_n=1.5\times 10^{-11}\, s$, $q=1$ and $\delta=0.09$.}
    \label{fig:ibrido_Dmu}
\end{figure*}

\begin{figure*}
    \centering
    \subfloat[]{
        \includegraphics[width=0.38\textwidth]{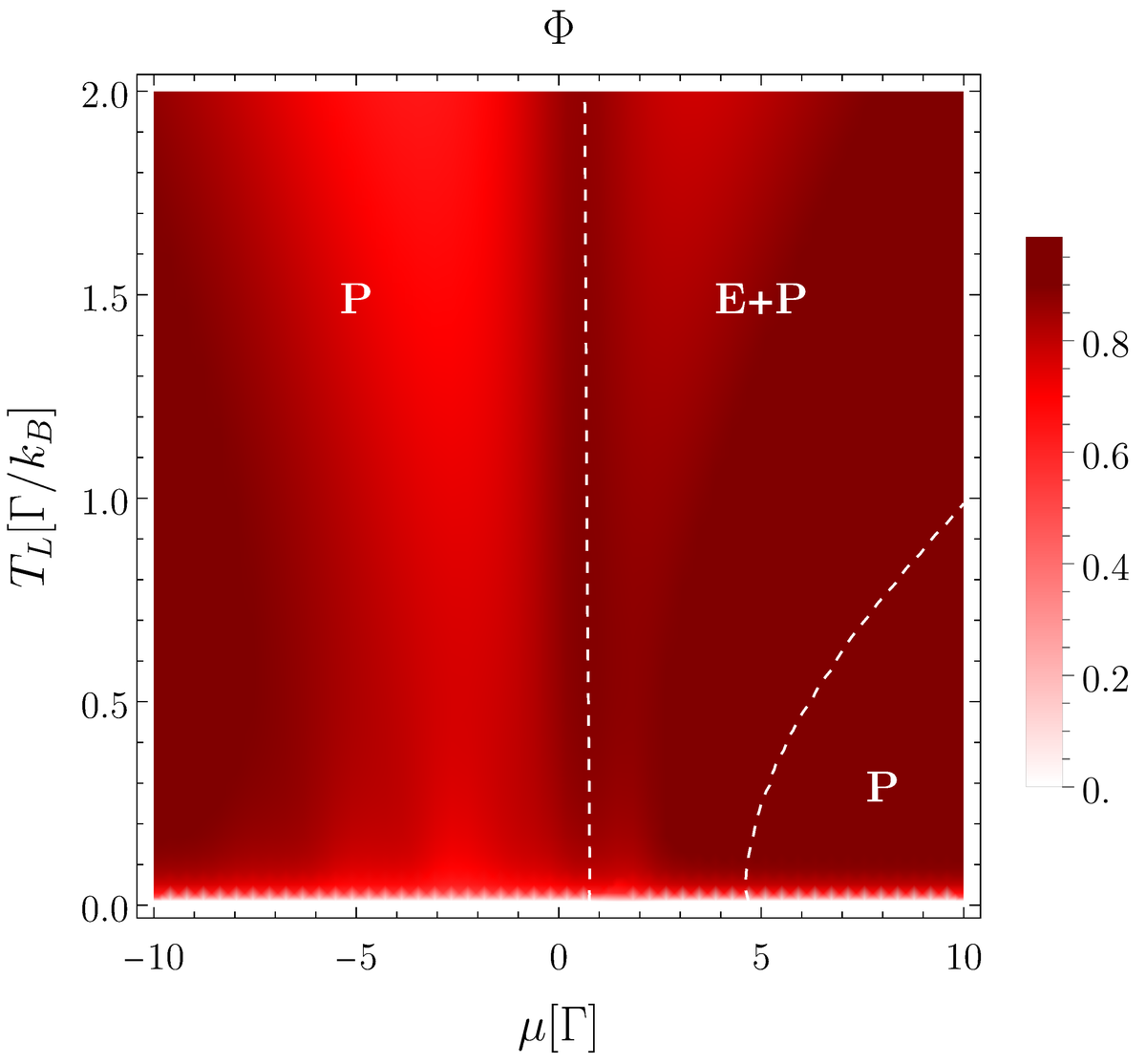}
        }
        \hspace{0.5cm}
    \subfloat[]{
        \includegraphics[width=0.38\textwidth]{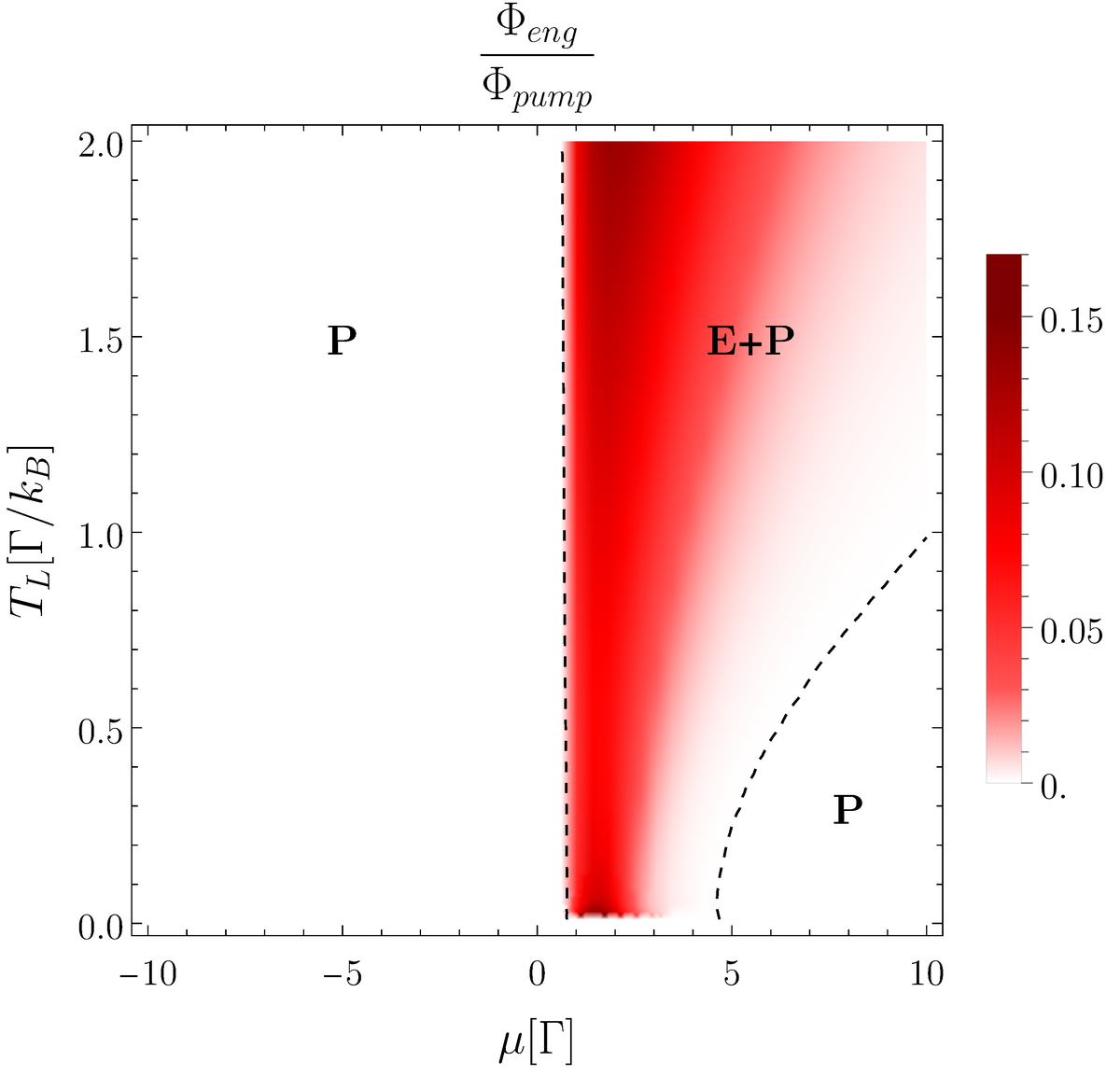}
        }
    \caption{(a) Density plot of exergy $\Phi$ as a function of $\mu$, in units of $\Gamma$, and $T_L$, in units of $\Gamma/k_B$. (b) Density plot of the ratio between the engine contribution and the heat pump contribution to the exergy, as a function of $\mu$, in units of $\Gamma$ and $T_L$, in units of $\Gamma/k_B$. In the regions marked with P the system acts as a heat pump, while regions where it operates as both a heat pump and an engine are marked with E+P. The considered parameters are: $T_R=0.01\,\Gamma/k_B$, $\Delta\mu=\Gamma$, $\Omega=2.6 \,\Gamma/\hbar$, $\tau_c=7.5\times 10^{-12}\, s$, $\tau_n=1.5\times 10^{-11}\, s$, $q=1$ and $\delta=0.09$.}
    \label{fig:ibrido_DT}
\end{figure*}
\end{widetext}

\section{Conclusions}
\label{Conclusions}

In this paper, we have studied a hybrid quantum thermal machine based on a small closed Hall channel at filling factor $\nu=1$, tunneling coupled to two terminals realized by quantum Hall states at $\nu=2$ and kept at different temperatures and chemical potentials. The left terminal is also driven out-of-equilibrium by means of the application of a periodic train of Lorentzian voltage pulses. Due to the presence of electron-electron interactions between the two edge channels at $\nu=2$ composing the terminals such pulses fractionalize. 

Considering the exergy as a figure of merit, we have determined interesting working regimes of this device such as the engine regime, where the power injected by the ac voltage is converted into useful electric power, a heat pump regime and also configurations where these two behaviors coexist. In addition, a refrigerator configuration can also be identified. Because in the present device electron-electron interactions between the channels can be treated exactly, we have also been able to characterize the robustness of the observed working regimes with respect to the interaction identifying regions in the parameter space where the efficiency is enhanced. 

Future developments in this work could address systems based on helical edge states of a two-dimensional topological insulator~\cite{Teo09, Strom09, Ferraro13}, fractional quantum Hall edge channels~\cite{Wen95, Kane95, Lee07, Carrega12} and more involved topological systems~\cite{Traverso24}, as well as multi-terminal configurations~\cite{Manzano20,Cavaliere23}.


\begin{acknowledgments}
Authors would like to thank R. Lopez, T. Martin, M. Moskalets, F. Ronetti and S. Ryu for useful discussions, and an anonymous referee for helpful criticism and suggestions. D.F. acknowledges the contribution
of the European Union-NextGenerationEU through the
``Quantum Busses for Coherent Energy Transfer'' (QUBERT) project, in the framework of the Curiosity Driven
2021 initiative of the University of Genova. G.B and D.F. acknowledge support from the project PRIN 2022 - 2022XK5CPX (PE3) SoS-QuBa - ``Solid State Quantum Batteries: Characterization and Optimization'' funded within the programme ``PNRR Missione 4 - Componente 2 - Investimento 1.1 Fondo per il Programma Nazionale di Ricerca e Progetti di Rilevante Interesse Nazionale (PRIN)'', funded by the European Union - Next Generation EU.
G.B. acknowledge support from the  Julian Schwinger Foundation (Grant JSF-21-04-0001) and  from
INFN through the project “QUANTUM”.

\end{acknowledgments}


\appendix

\section{Derivation of $I^{u}_{L}$ in Eq.~(\ref{IuL})}
\label{App1}
We start from the general expression in Eq.~(\ref{I^u}) of the main text)
\begin{equation}\label{Iu_cons}
    I_{L}^u=\int_{-\infty}^{+\infty}\frac{dE}{2 \pi \hbar}E[f_{L}^{out}(E)-f_{L}(E)].
\end{equation}
Taking into account the constraints in Eq.~(\ref{constraints}) and the defintion of $f^{out}_{L}(E)$ in Eq.~(\ref{f_j_out}) one can decompose it as 
\begin{equation}
I_{L}^u=I_{L,1}^u+I_{L,2}^u
\end{equation}
with 
\begin{eqnarray}
I_{L,1}^u&=&\int_{-\infty}^{+\infty}\frac{dE}{2 \pi \hbar}E|S_{LR}(E)|^{2}\left[f_{R}(E)-\sum_{n=-\infty}^{+\infty}|\mathcal{P}_{-n}|^{2}f_{L}(E_{n})\right]\nonumber\\
I_{L,2}^u&=&\int_{-\infty}^{+\infty}\frac{dE}{2 \pi \hbar}E\left[\sum_{n=-\infty}^{+\infty}|\mathcal{P}_{-n}|^{2}f_{L}(E_{n})-f_{L}(E)\right].\nonumber\\
\end{eqnarray}
These two terms need to be discussed separately. 

Indeed, the former can be rewritten through proper shifts of the second term in energy domain ($E\rightarrow E-n \hbar \Omega$ and $n\rightarrow -n$) in the form 
\begin{eqnarray}
I_{L,1}^u&=&\int_{-\infty}^{+\infty} \frac{dE}{2 \pi \hbar} \left[E |S_{LR}(E)|^{2} f_R(E)+\right.\nonumber\\
&-& \left.\sum^{+\infty}_{n=-\infty}|\mathcal{P}_{n}|^{2}\left(E+n \hbar \Omega\right)|S_{RL}(E_{n})|^{2}f_L(E)\right]
\\
&=&\int_{-\infty}^{+\infty} \frac{dE}{2 \pi \hbar}E[\tau_{LR}(E)f_R(E)-\tau_{RL}(E)f_L(E)]+\nonumber\\
&+&\frac{\Omega}{2\pi}\sum_{k=L,R}\int_{-\infty}^{+\infty} dE\langle n(E)\rangle_{Lk}f_k(E),\\
\end{eqnarray}
where we have taken into account the symmetries in Eq.~(\ref{constraints}) and in the second line we have used the definitions in Eqs.~(\ref{TRL}), (\ref{TLR}) and (\ref{n_medio}). 

The $I^{u}_{L,2}$ term is more pathological, being the difference of two separately divergent terms and need to be treated carefully. To do so, we proceed by Taylor expanding the frequency dependent part in such a way that 

\begin{equation}
I_{L,2}^u=\int_{-\infty}^{+\infty}\frac{dE}{2 \pi \hbar}E\left[\sum_{n=-\infty}^{+\infty}|\mathcal{P}_{-n}|^{2}\sum^{+\infty}_{s=0}\frac{d^{s} f_{L}(E)}{d E^{s}} \frac{(\hbar n \Omega)^{s}}{s!}-f_{L}(E)\right].
\end{equation}
Taking into account the general relation~\cite{Vannucci17}
\begin{equation}
\sum^{+\infty}_{n=-\infty}n^{s} |\mathcal{P}_{n}|^{2}= \left(\frac{e}{\hbar \Omega}\right)^{s}\frac{1}{\mathcal{T}}\int^{+\frac{\mathcal{T}}{2}}_{-\frac{\mathcal{T}}{2}} V^{s}_{1, out}(t) dt 
\end{equation}
one has that the $s=0$ term cancels out because $\sum^{+\infty}_{n=-\infty}|\mathcal{P}_{n}|^{2}=1$, the $s=1$ term is zero due to the purely ac nature of the incoming drive ($\sum^{+\infty}_{n=-\infty}n|\mathcal{P}_{n}|^{2}=0$). Moreover all the terms $s>2$ are zero due to the fast decaying of the derivatives of the Fermi distribution at energies $\pm \infty$. Therefore the only remaining contribution is the one with $s=2$ leading to 
\begin{equation}
I_{L,2}^u=P_{in}=\frac{e^{2}}{4 \pi \hbar}\frac{1}{\mathcal{T}}\int^{+\frac{\mathcal{T}}{2}}_{-\frac{\mathcal{T}}{2}} V^{2}_{1, out}(t) dt.
\end{equation}
This justifies the result reported in Eq.~(\ref{P_in}). 

Two comments are in order at this point. First of all, the latter voltage dependent contribution is not present in $I^{u}_{R}$ (see Eq.~(\ref{IuR})). This is due to the fact that no ac drive is applied directly to the $R$ terminal. Moreover, carrying out the same calculation for the charge current $I^{e}_{L}$ defined in Eq.~(\ref{charge_current}) we can identify also there a potentially pathological term. However, it is proportional to $\sum^{+\infty}_{n=-\infty}n|\mathcal{P}_{n}|^{2}=0$. This justifies the expressions reported in the main text in Eq.~(\ref{IeL}) for the charge currents.

\section{Supplementary plots of the relevant physical quantities}
\label{App2}

\subsection{Plot of $P_{e}$ in the engine regime}

Here, we report the density plots of $P_{e}$ at $T_{L}=T_{R}$ as a function of $\mu$ and $\Delta \mu$. We present only the regions where $P_{e}$ is positive (red regions of the density plot). Figure \ref{fig:Pe_engine} (a) represents the non-interacting case ($\theta=0$), while Figure \ref{fig:Pe_engine} (b) shows the strongly interacting case ($\theta=\pi/4$). In full agreement with what was discussed in the main text, we observe that the behavior is qualitatively very similar in the two cases. 

\begin{widetext}
\begin{figure*}[h]
    \centering
    \subfloat[]{
        \includegraphics[width=0.38\textwidth]{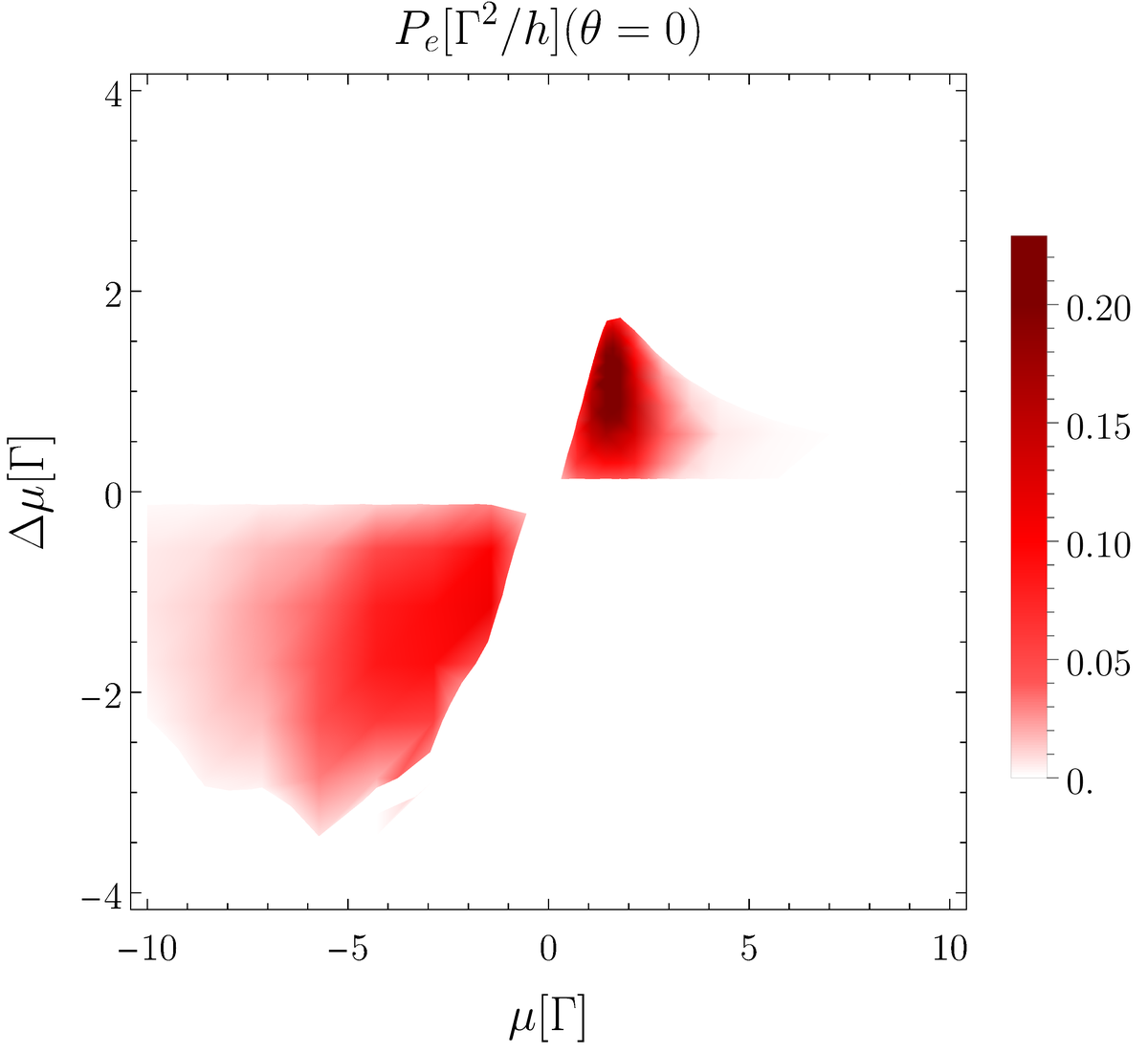}
        }
        \hspace{0.5cm}
    \subfloat[]{
        \includegraphics[width=0.38\textwidth]{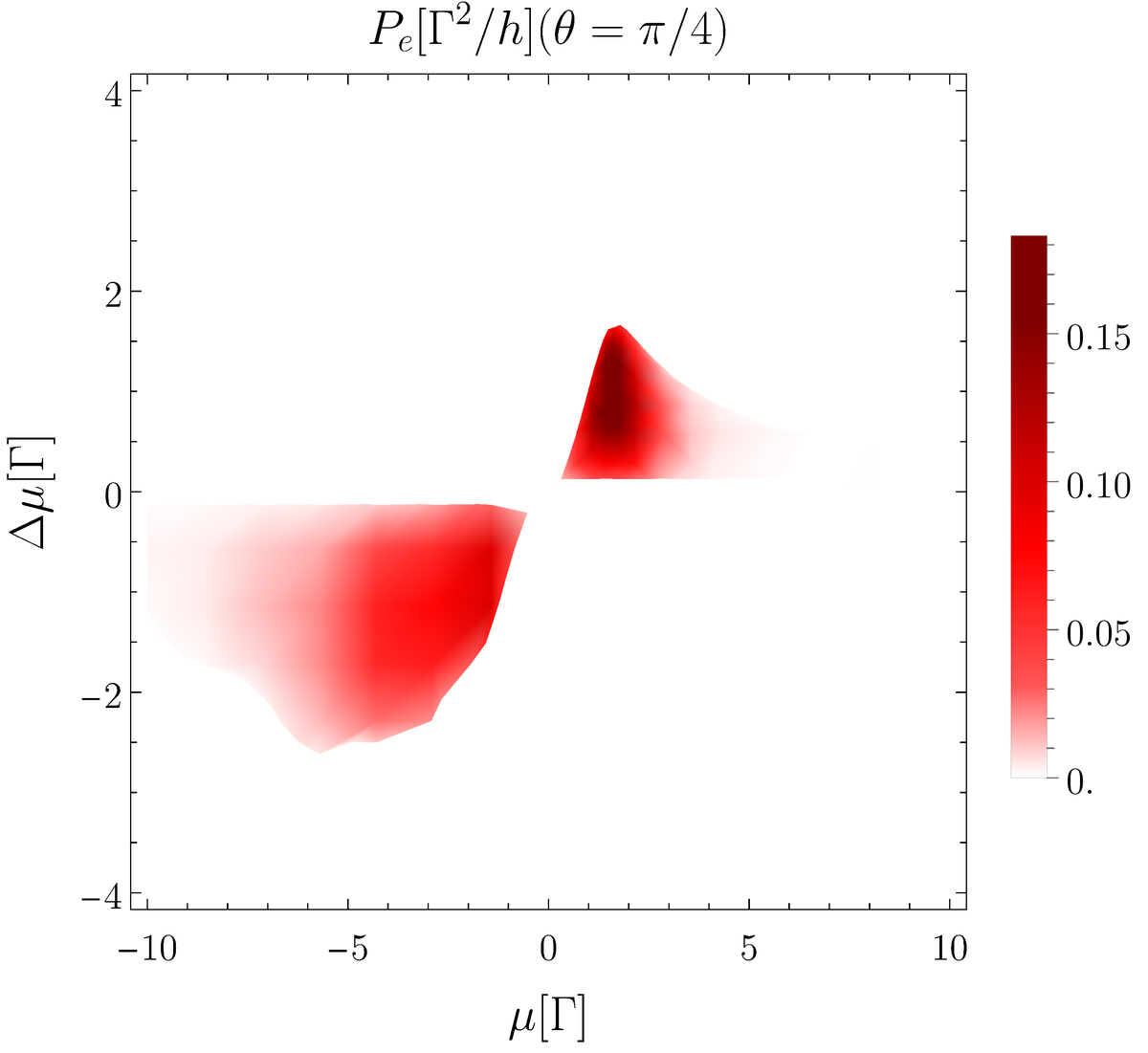}
        }
    \caption{Density plots of $P_e$ (in units of $\Gamma^2/\hbar$) as a function of $\mu$ and $\Delta\mu$ (in units of $\Gamma$), in the non-interacting case (for $\theta=0$), in panel (a), and with interactions ($\theta=\pi/4$), in panel (b). The other parameters are: $T_R=T_L=0.01\,\Gamma/k_B$, $\Omega=2.6 \,\Gamma/\hbar$, $\tau_c=7.5\times 10^{-12}\, s$, $\tau_n=1.5\times 10^{-11}\, s$, $q=1$ and $\delta=0.09$.}
    \label{fig:Pe_engine}
\end{figure*}

\begin{figure*}
    \centering
    \subfloat[]{
        \includegraphics[width=0.38\textwidth]{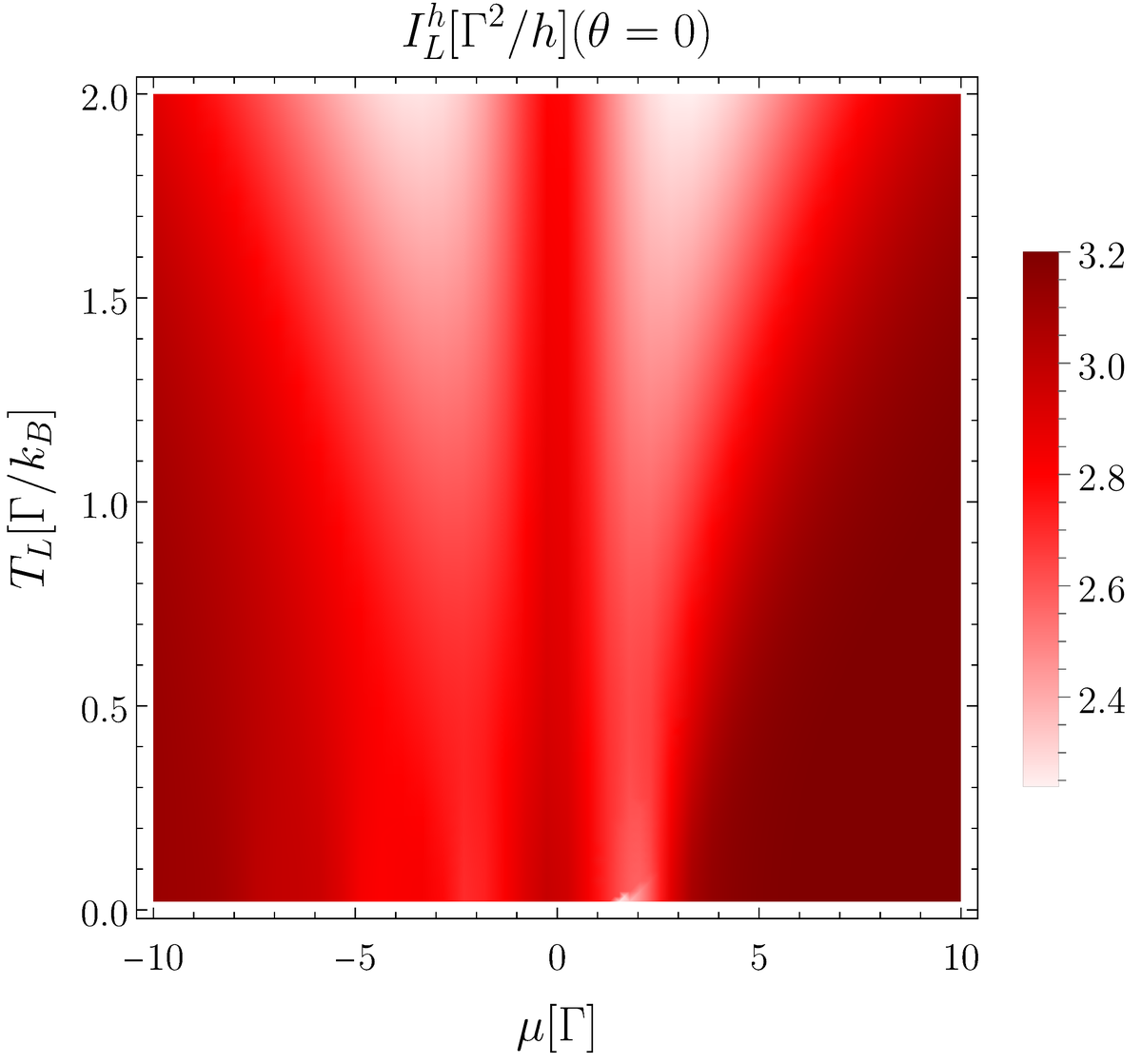}
        }
        \hspace{0.5cm}
    \subfloat[]{
        \includegraphics[width=0.38\textwidth]{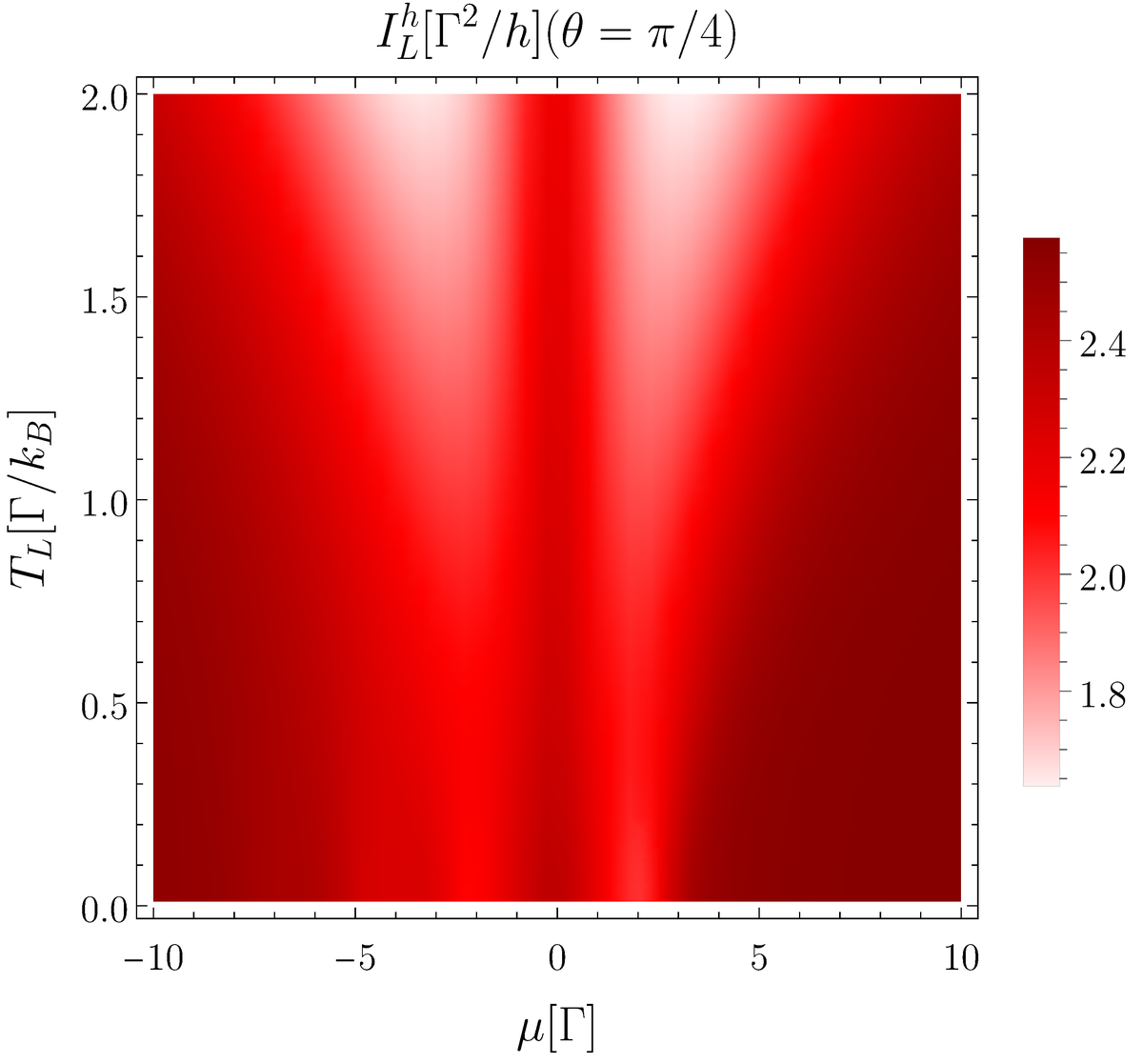}
        }
    \caption{Density plots of $I_L^h$ (in units of $\Gamma^2/\hbar$) as a function of $\mu$ (in units of $\Gamma$) and $T_L$ (in units of $\Gamma/k_B$) in the non-interacting case (for $\theta=0$), in panel (a), and with interactions ($\theta=\pi/4$), in panel (b). The other parameters are: $T_R=0.01\,\Gamma/k_B$, $\Delta\mu=\Gamma$, $\Omega=2.6 \,\Gamma/\hbar$, $\tau_c=7.5\times 10^{-12}\, s$, $\tau_n=1.5\times 10^{-11}\, s$, $q=1$ and $\delta=0.09$.}
    \label{fig:IhL_pump}
\end{figure*}
\end{widetext}

\subsection{Plot of $I^{h}_{L}$ in the pump regime}

Proceeding in analogy with what was done for $P_{e}$, we show here the density plot of $I^{h}_{L}$ in the $\mu$, $T_{L}$ plane at fixed $\Delta \mu$ and $T_{R}$. Also in this case is possible to observe the robustness of the phenomenology discussed with respect to the electron-electron interaction in the quantum Hall channels (compare the left and right panels of Figure \ref{fig:IhL_pump}). Notice that, differently from what happens for $P_{e}$, here $I^{h}_{L}$ is always positive.

\section{Considerations about the drive frequency $\Omega$}
\label{App3}

\begin{widetext}
\begin{figure*}
    \centering
    \subfloat[]{
        \includegraphics[width=0.35\textwidth]{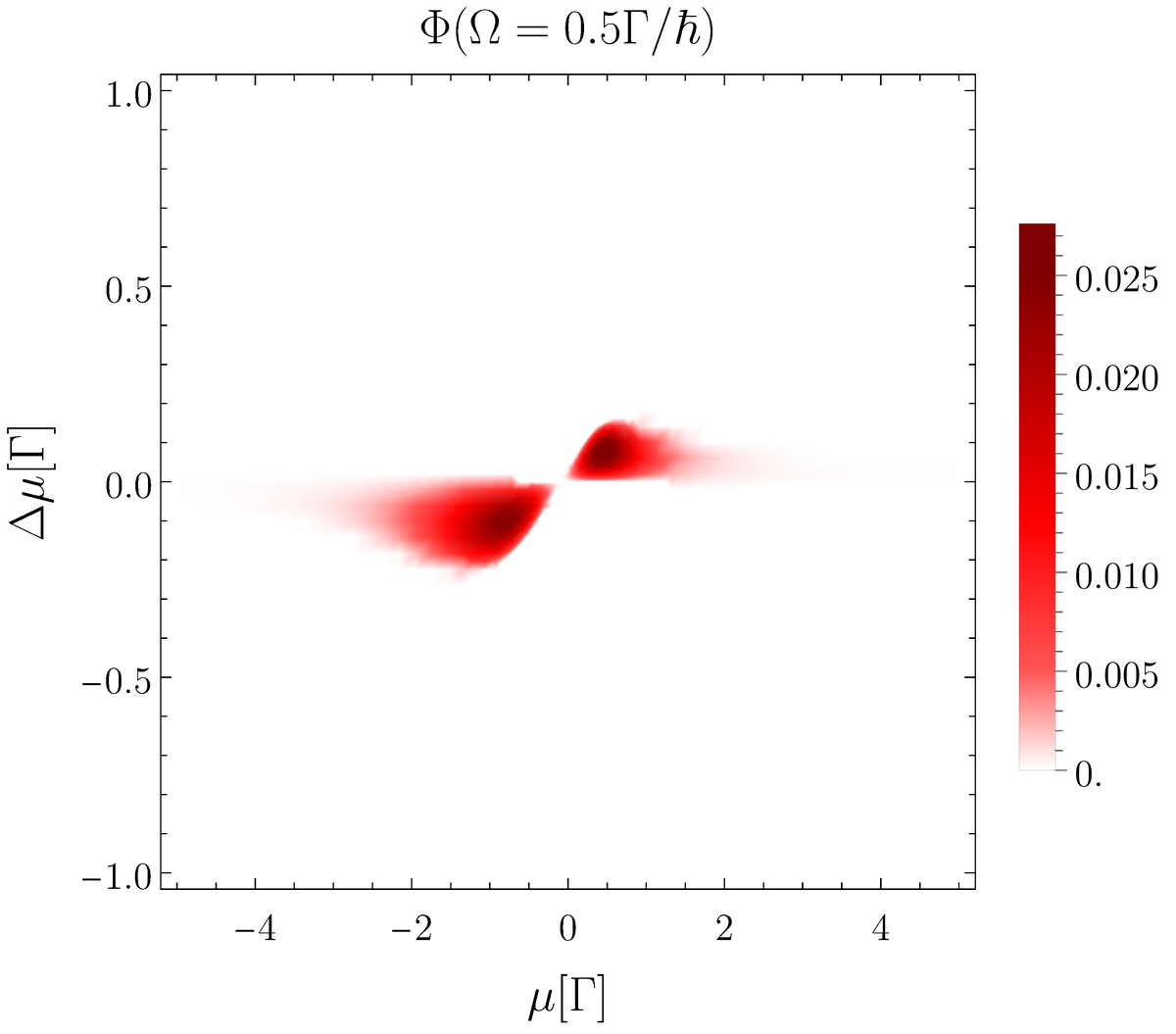}
        }
        \hspace{0.5cm}
    \subfloat[]{
        \includegraphics[width=0.35\textwidth]{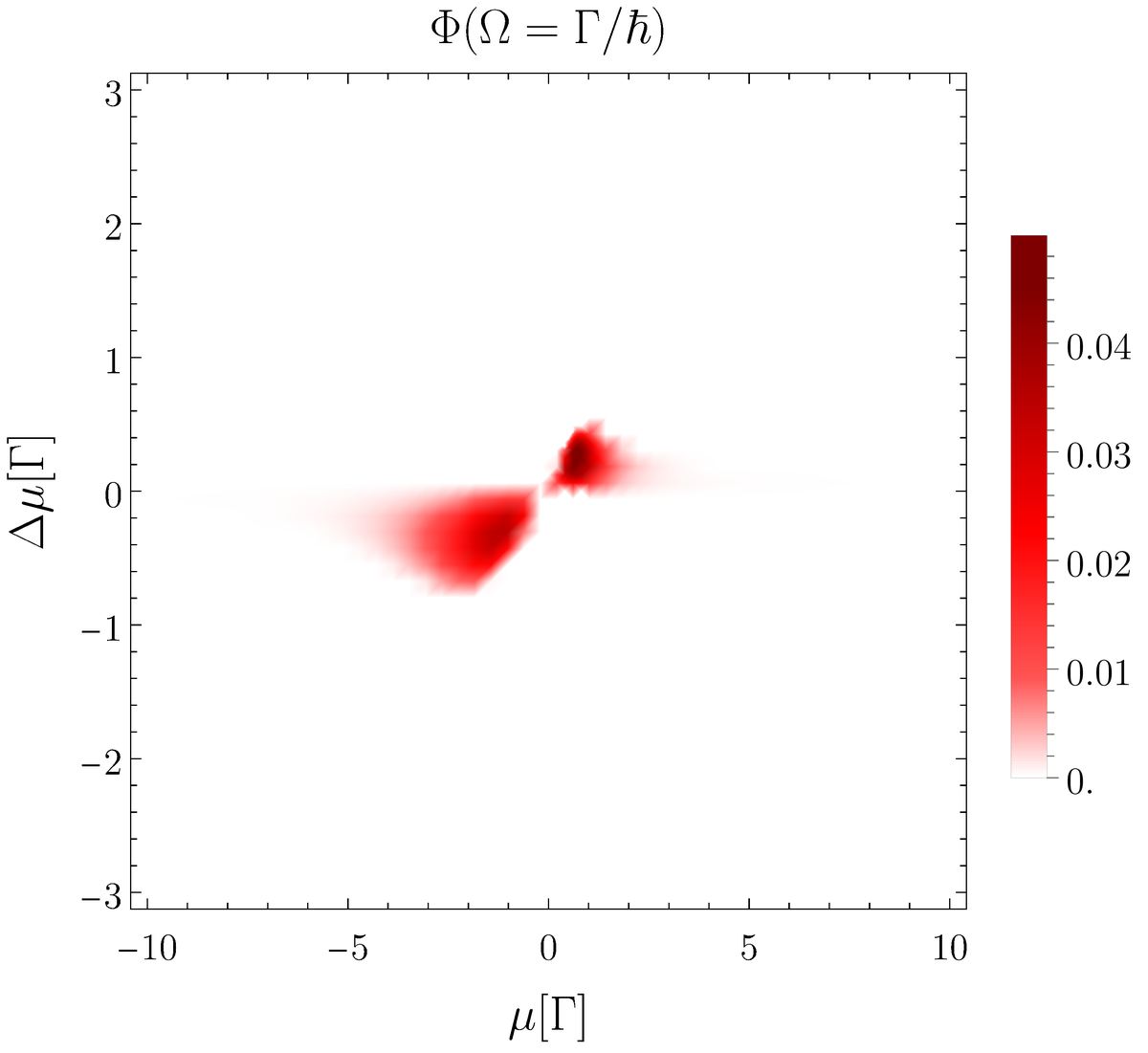}
        }\\
    \subfloat[]{
        \includegraphics[width=0.35\textwidth]{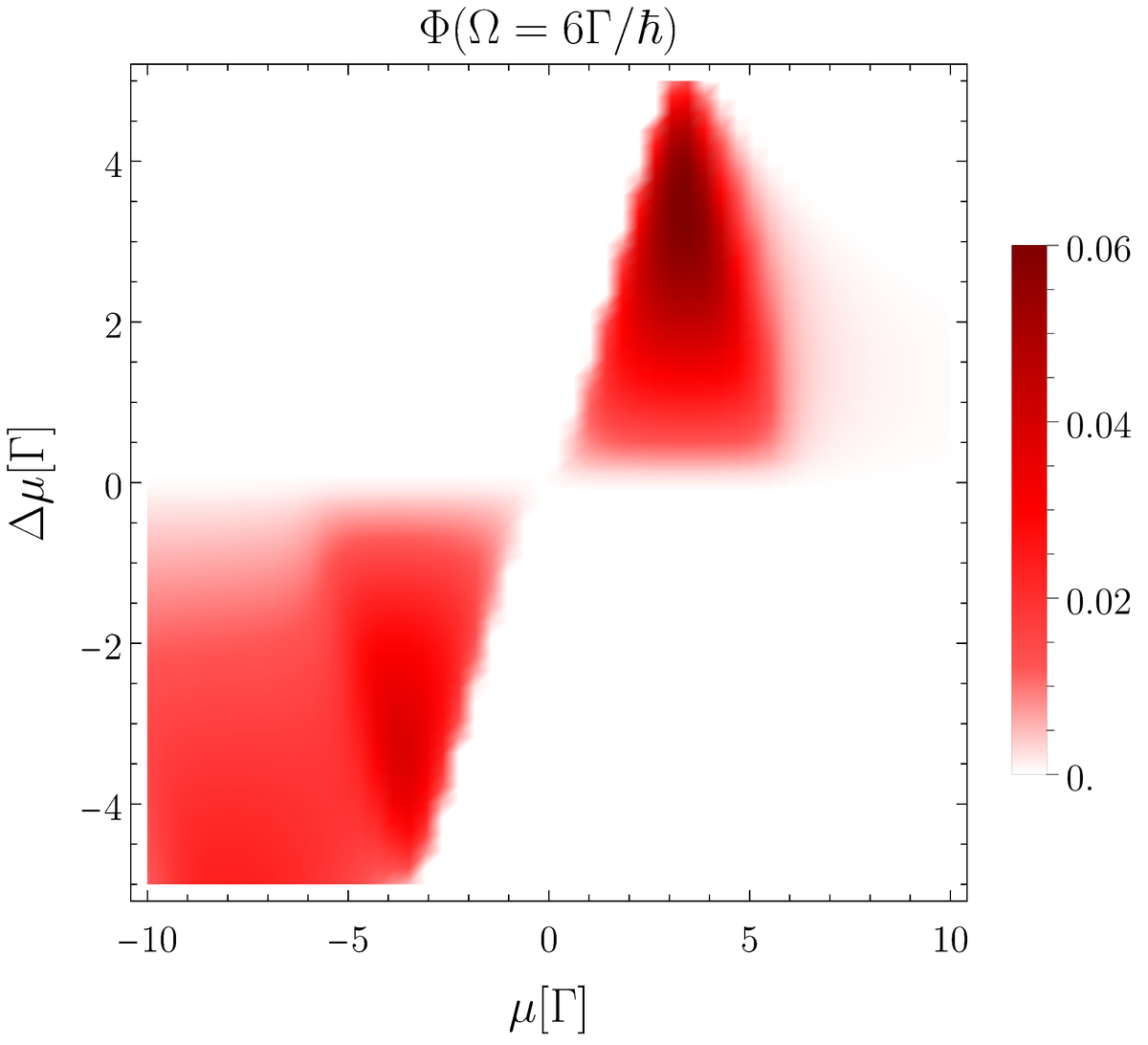}
        }
        \hspace{0.5cm}
    \subfloat[]{
        \includegraphics[width=0.35\textwidth]{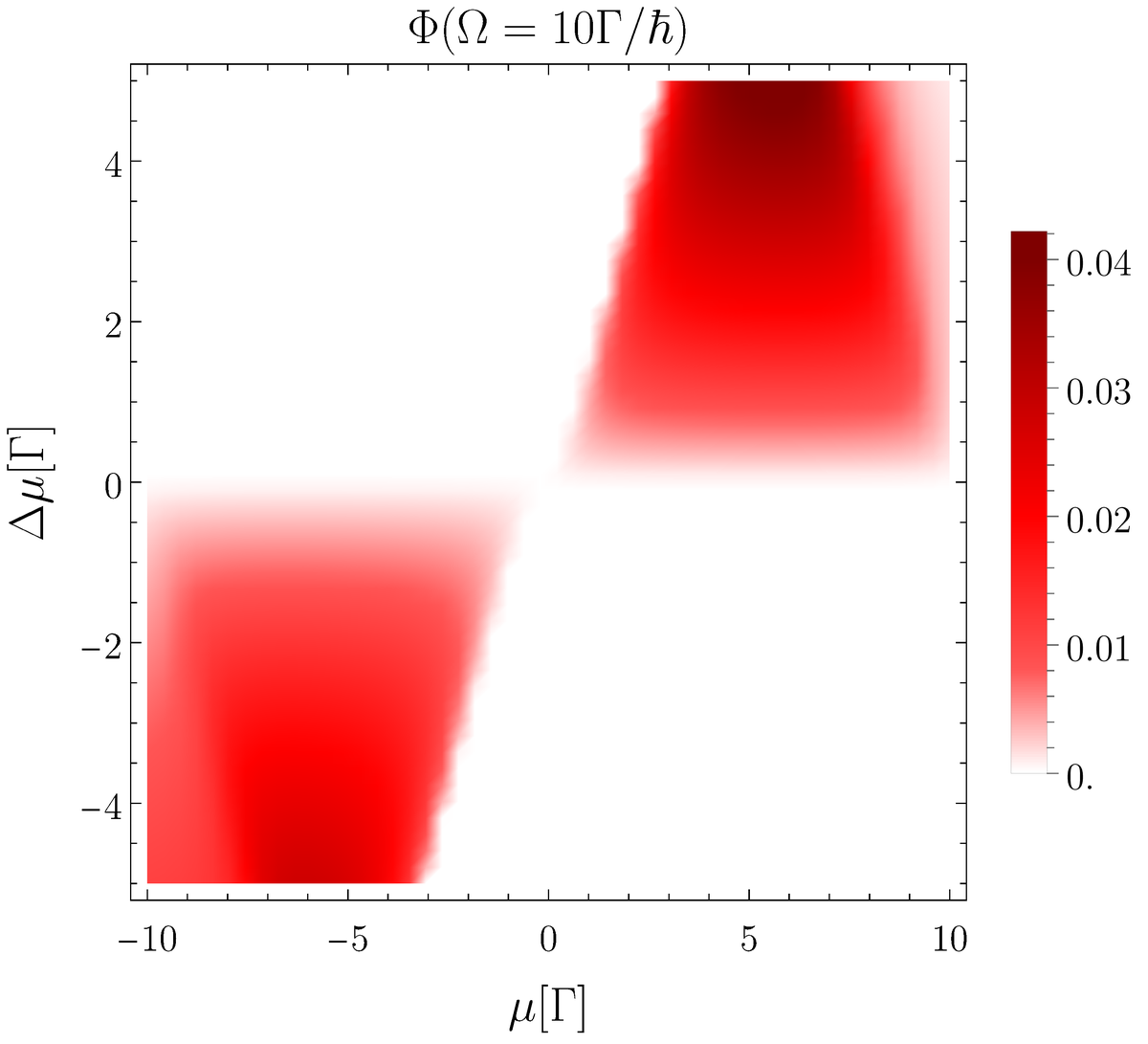}
        }
    \caption{Density plots of exergy $\Phi$ in the engine regime as a function of $\mu$ and $\Delta \mu$ (in units of $\Gamma$), for different values of $\Omega$. The considered parameters are: $\theta=0$, $T_R=T_L=0.01 \,\Gamma/k_B$, $\tau_c=7.5\times 10^{-12}\, s$, $\tau_n=1.5\times 10^{-11}\, s$, $q=1$ and $\delta=0.09$.}
    \label{fig:Phi_Omega}
\end{figure*}
\end{widetext}

Here, we discuss the role of the drive frequency in the behavior of $\Phi$ in the engine regime. In Figure \ref{fig:Phi_Omega} we observe that in the non-interacting case at low frequencies $\hbar \Omega \lesssim \Gamma $ the region where the considered device works as an engine is considerably reduced with respect to what was discussed in the main text (see panels (a) and (b)). Moreover, in this regime, one has an exergy $\Phi\lesssim 4\%$. By increasing the frequency ($\hbar \Omega \gg \Gamma $) the positive power region $P_{e}$ is larger, but the exergy remains below $\Phi\approx 6\%$ (see panels (c) and (d)). This indicates that a good compromise between the effective region of engine behavior and the efficiency of this work-to-work conversion is achieved for an intermediate frequency value such as the one considered in the main text ($\hbar \Omega=2.6 \Gamma$).

\section{Refrigerator regime}
\label{App4}

The proposed device also shows a refrigerator behavior which is properly described in terms of the heat current $I^{h}_{R}$ extracted from the cold right lead ($T_{R}<T_{L}$). According to this the exergy is defined as 

\begin{equation}
\Phi=\frac{-I^{h}_{R}\Theta(-I^{h}_{R})\left(\frac{T_{L}}{T_{R}}-1\right)+P_{e}\Theta(P_{e})}{P_{in}+I^{h}_{R}\Theta(I^{h}_{R})\left(\frac{T_{L}}{T_{R}}-1\right)-P_{e}\Theta(-P_{e})}.
\end{equation}

\begin{widetext}
\begin{figure*}[h]
    \centering
    \subfloat[]{
        \includegraphics[width=0.38\textwidth]{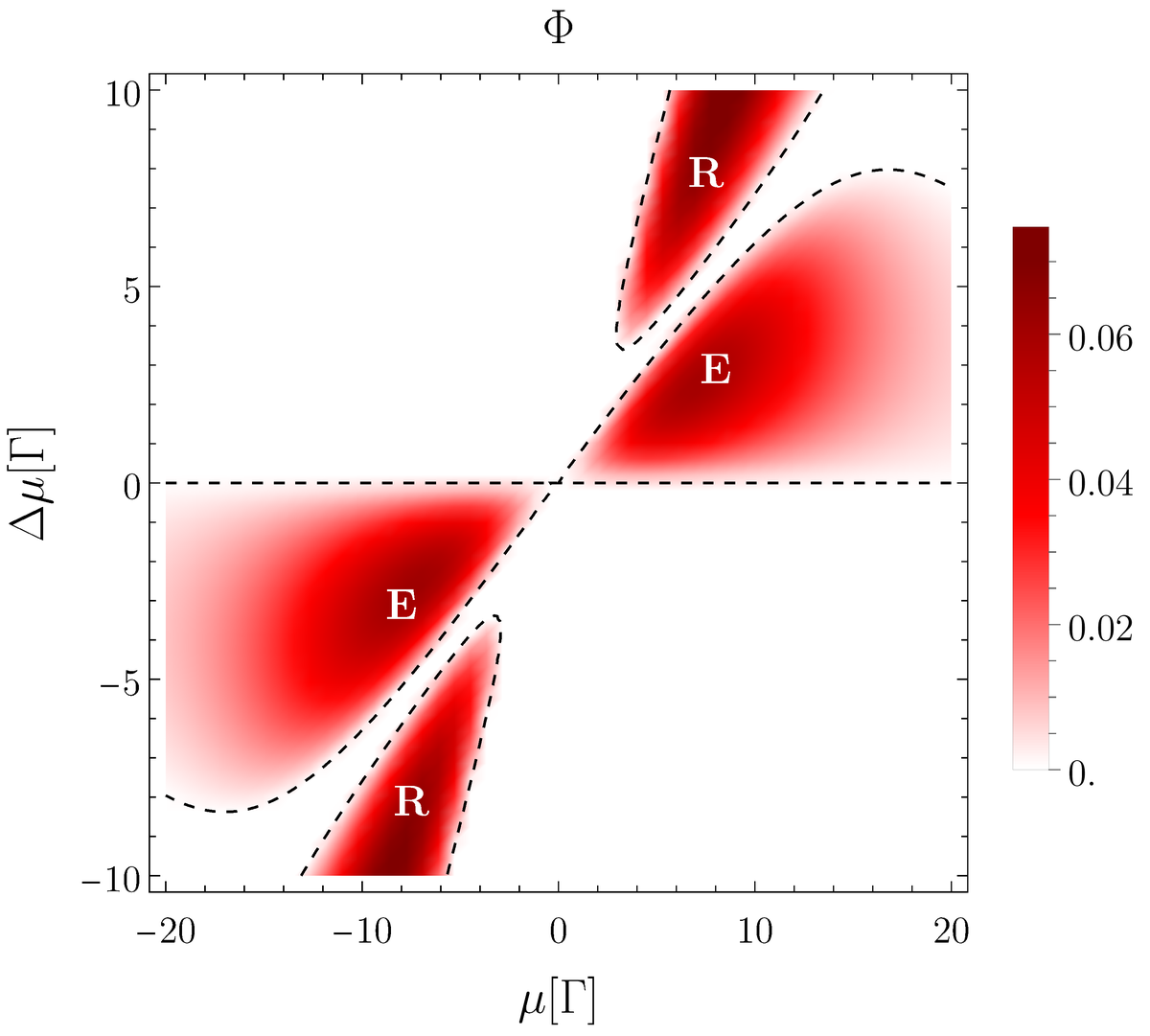}
        }
        \hspace{0.5cm}
    \subfloat[]{
        \includegraphics[width=0.38\textwidth]{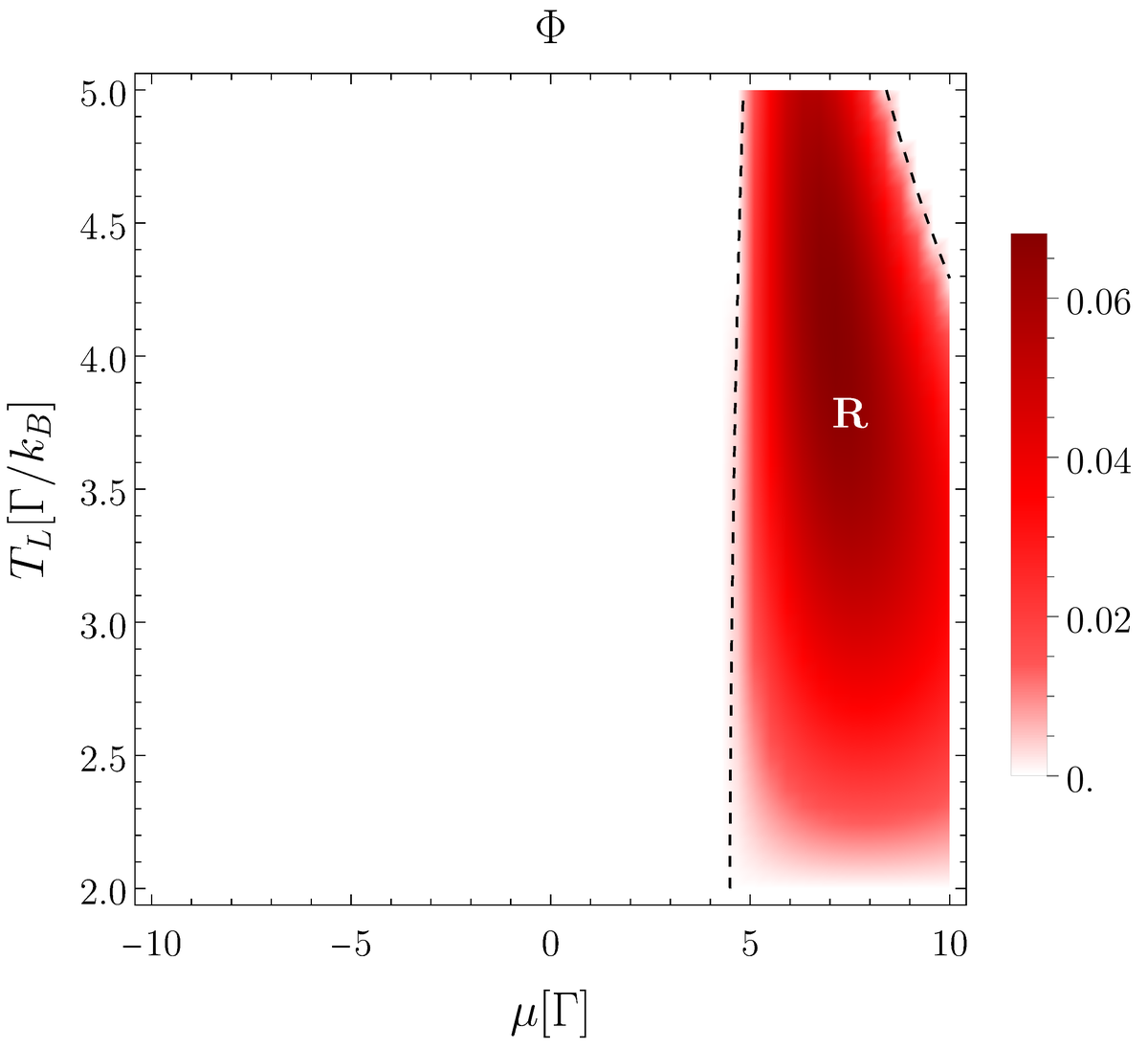}
        }
    \caption{(a) Density plot of exergy $\Phi$ as a function of $\mu$ and $\Delta \mu$ (in units of $\Gamma$), for $T_R=2\Gamma/k_B$ and $T_L=4\Gamma/K_B$. (b) Density plot of exergy $\Phi$ as a function of $\mu$, in units of $\Gamma$, and $T_L$, in units of $\Gamma/k_B$, with $\Delta\mu=8\Gamma$. The system acts as a refrigerator in the regions marked with R, and as an engine in the regions marked with E. The other parameters are: $\Omega=2.6 \,\Gamma/\hbar$, $\tau_c=7.5\times 10^{-12}\, s$, $\tau_n=1.5\times 10^{-11}\, s$, $q=1$ and $\delta=0.09$.}
    \label{fig:Phi_frigo}
\end{figure*}
\end{widetext}

Proceeding as in the main text, it is possible to determine the regions in the $\mu$, $\Delta \mu$ plane and at the fixed $\Delta T$ where the refrigerator and engine behavior emerge. However, in this case there is no overlap between the two working regimes (see Figure \ref{fig:Phi_frigo} (a)). As a function of $\mu$ and $T_{L}$ and at fixed $\Delta \mu$ the refrigerator regime can be achieved at quite high averaged chemical potentials $\mu \gtrsim 4.5 \Gamma $. In both the discussed cases the exergy approaches at most $\Phi\approx 7\%$.  

\clearpage
\bibliography{QPC}

\end{document}